\documentclass[%
 aip,
 amsmath,amssymb,
 reprint,%
]{revtex4-1}

\usepackage{graphicx}
\usepackage{dcolumn}
\usepackage{bm}
\usepackage[marginal]{footmisc}
\usepackage[utf8]{inputenc}
\usepackage[T1]{fontenc}
\usepackage{mathptmx}
\usepackage{graphicx,subfigure}
\usepackage{float}
\usepackage{color}
\usepackage{CJKutf8}

\begin{document}

\preprint{AIP/123-QED}

\title{\textcolor{black}{Instability and Mixing of Gas Interfaces Driven by Cylindrically Converging Shock Wave}}


\author{\textcolor{black}{Wei-Gang Zeng}} 

\affiliation{ 
Department of Engineering Mechanics, School of Aerospace, Tsinghua University, Beijing 100084, China.}%

\author{\textcolor{black}{Yu-Xin Ren }}
\affiliation{ 
Department of Engineering Mechanics, School of Aerospace, Tsinghua University, Beijing 100084, China.}%

\author{\textcolor{black}{Jianhua Pan}}
\email{pan.796@osu.edu}
\affiliation{
Department of Engineering Mechanics, School of Aerospace, Tsinghua University, Beijing 100084, China.}%
\affiliation{ 
William. G. Lowrie Department
of Chemical and Biomolecular Engineering,
Koffolt Labs, The Ohio State University, US.}%

\date{\today}

\begin{abstract}
In the present paper, an efficient method to generate "pure" cylindrically converging shock wave without a following contact surface is proposed firstly. Then, the Richtmyer-Meshkov instabilities of two interfaces driven by the generated cylindrically converging shock wave and the associated fluids’ mixing behaviors are numerically studied. The results show that the instability of the interface is characterized by the growth of perturbation amplitude before re-shock. However, the mixing of fluids is enhanced dramatically after re-shock, which is manifested not only by the evolutions of flow structures but also by the temporal behaviors of mixing parameters. Further investigation shows that, although these two cases are of different initial perturbations, their evolutions of mixing width and other mixing parameters such as  molecular mixing fraction, local anisotropy and density-specific volume correlation could achieve the same laws of temporal behavior, especially during the later stage after re-shock. These results to some extent demonstrate that there also exist scaling law and temporal asymptotic behaviors in the mixing zone for cylindrically converging shock wave driven interface.  Moreover, the analyses of turbulent kinetic energy spectrums in the azimuthal direction at late stage also witness the $k^{-5/3}$ decaying law of turbulent kinetic energy for the present inhomogeneity flows driven by cylindrically converging shock wave, which further manifests that the fluids’ mixing is indeed enhanced at later time after re-shock.
\end{abstract}

\maketitle

\section{\label{sec:level1} INTRODUCTION}
The interaction between shock and gases interface with initial perturbations is complicated, which involves interfacial instabilities of various kinds and the possible turbulent mixing at late stage. At the beginning of interaction, the Richtmyer-Meshkov instability (RMI) dominates the flow, resulting in the deposition of baroclinic vorticities and the growth of the perturbation. Then, the accumulated vortices cause the formation of the primary Kelvin–Helmholtz (KH) billows, which would lead to the dramatic mixing of fluids when the more complex instabilities occur in these billows.

In the last three decades, the instabilities of shock driven gases interface and the associated fluids’ mixing behaviors have been widely studied due to their importance in Inertial Confinement Fusion (ICF) \cite{lindl1995development}, supernova explosions \cite{arnett2000role}, and supersonic combustion \cite{yang1993applications}. Many theoretical models are proposed for the vortex generation \cite{picone_boris_1988, yang_kubota_zukoski_1994, samtaney_zabusky_1994, niederhaus_greenough_oakley_ranjan_anderson_bonazza_2008} and growth rate of perturbation amplitude \cite{ZhangNonlinear_1997, VandenboomgaerdeNonlinear_2002, MatsuokaNonlinear_2003, SohnSimple_2003} of RMI flows, \textcolor{black}{most of which are manifested not only by numerical simulations \cite{samtaney_zabusky_1994, niederhaus_greenough_oakley_ranjan_anderson_bonazza_2008, li2019circulation, wang2020numerical, sun2020microscopic, liang2020interfacial} but also by experiments \cite{wang2015interaction, liu2018elaborate, zhai2019coupling}.} Recently, the essential development of fluids’ mixing at late stage attracts great attention \textcolor{black}{\cite{zhou2017rayleigh}}. Therefore, the morphological behaviors of turbulent mixing and the criterion of turbulent mixing transition are extensively studied during past ten years  \textcolor{black}{\cite{Hill2006Large, Thornber2010The, Tritschler2014On, Lombardini2012Transition, Thornber2017Late, Olson2014Comparison, thornber2019turbulent, zhou2019turbulent, li2019role}}. However, most of these investigations are mainly focused on interface instabilities and the associated fluids’ mixing induced by planar shock wave. By contrast, the studies on interface instabilities and the associated fluids’ mixing induced by converging shock wave, which would be more relevant to engineering applications such as ICF, are much less abundant. \textcolor{black}{Obviously, these research topics are gaining increasing attention recently. For example, recent studies have manifested that, due to the Bell-Plesset effect and Rayleigh-Taylor effect, the perturbation amplitude of interface driven by converging shock wave would grow in a different way both at early and later stages \cite{mikaelian2005rayleigh, ding2017measurement, liu2019richtmyer, luo2019nonlinear, suponitsky2014richtmyer, Rafei2019three}. However, the fluids’ mixing behaviors of interface driven by converging shock wave at late stage remain to be open issues.} 


The motivation of the present paper is to investigate the instability and the associated fluids’ mixing of gases interfaces driven by cylindrically converging shock wave (CCSW) using high resolution finite volume (FV) method. To this end, we adopt an efficient way to generate “pure” CCSW which can be used as the incident shock wave for further numerical study of interaction between CCSW and gases interface. Then, the RMI and the associated fluids’ mixing behaviors of two interfaces driven by such CCSW are numerically investigated in detail. The evolutions of flow structure and fluids’ mixing behaviors highlight that the instability of the interface is characterized by the growth of perturbation amplitude before re-shock, while the fluids’ mixing is enhanced dramatically after the interface being re-shocked. The enhanced fluids’ mixing is also manifested by the exponential scaling laws of mixing width as well as the temporal asymptotic behaviors of mixing parameters such as molecular mixing fraction, local anisotropy and density-specific volume correlation at later stage after re-shock. Moreover, the turbulent kinetic (TKE) in the azimuthal direction also decays with a slop of $k^{-5/3}$ in a relatively broader range of low wave numbers at later stage after re-shock, which further confirms the enhanced fluids’ mixing since the inertial range is extended during the developing process. 

The remainder of this paper is organized as follows. The numerical framework based on high resolution FV method for compressible two fluids is presented in Section II. The method to generate “pure” CCSW and the verification for its usage in CCSW/interface interaction are introduced in Section III. The numerical simulations for the CCSW induced RMI flows and the corresponding results which include the wave patterns, the fluids' mixing behaviors and the decaying law of TKE spectrums are well discussed in Section IV. And finally, the conclusions remarks are given in Section V.

\section{NUMERICAL FRAMEWORK}
\subsection{Governing equations}
Following our previous work\cite{zeng2018numerical, zeng2018turbulent, wang2020consistent}, the integral form of governing equations for compressible two fluids with consistent treatment of the convective terms at a material interface can be written as
\begin{equation}
\frac{\partial }{\partial t}\int{\bm Q}d{\Omega}=\oint_{\partial\Omega}(\bm{F}_c-\bm{F}_v)\cdot\bm{n}dS=\int{\bm W}d{\Omega}\ .
\end{equation}
In the above formula, $\bm{Q}=\begin{bmatrix}\rho &\rho Y_i & \rho \bm{u} & \rho E & \theta\end{bmatrix}^T $ is the vector of quasi-conservative variables in the control volume $\Omega$, where $\rho$ is the density of mixture, $Y_i$ is the mass fraction of specie $i$, $\bm{u}=[u\ v\ w]^T$ is the vector flow velocity, $E$ is the total energy of mixture and $\theta=\dfrac{1}{\gamma-1}$ is the function of the specific heat ratio of mixture, $\gamma$. It should be noted that, due to the quasi-conservative form of $\theta$, a source term $\bm{W}= \begin{bmatrix}0 &0 &0 &0 & \theta \nabla{\cdot{\bm{u}}}\end{bmatrix}^T$ should be added to the right-hand-side of Eq. (1). In fact, the equation for $\theta$ is introduced to achieve a consistent treatment of the material interfaces and to remove non-physical oscillations in the vicinity of the material interfaces, which was initially proposed by Abgrall \cite{abgrall1996prevent} and lately improved by Johnsen \cite{johnsen2012preventing} for FV method using high order reconstructions. Additionally, $\bm{F}_c$ and $\bm{F}_v$ are, respectively, the inviscid flux and viscous flux on the control surface $S$ with the unit outward normal vector $\bm{n}$. Their definitions are given as follows
\begin{equation}
\bm{F}_c=\begin{bmatrix}
\rho \bm{u}\\ 
\rho Y_i \bm{u}\\ 
\rho \bm{u} \bm{u}+p\textbf{I}\\ 
\rho H \bm{u}\\ 
\theta \bm{u}
\end{bmatrix}, 
\end{equation}

\begin{equation}
\bm{F}_v=\begin{bmatrix}
0 \\ 
\rho D_i \nabla{Y_i}\\ 
\underline{\bm{\tau}}\\ 
\underline{\bm{\tau}}\cdot \bm{u}-\bm{q}_c-\bm{q}_d\\ 
0
\end{bmatrix}.
\end{equation}
In the above two equations, $\textbf{I}$ is the unit tensor, $H=E+p/\rho$ is the total enthalpy of the mixture, $\underline{\bm{\tau}}=2\mu\underline{\bm{S}}-\dfrac{2}{3}\mu (\nabla{\cdot{\bm{u}}}) \textbf{I}$ is the viscous stress tensor, and $\underline{\bm{S}}=\dfrac{1}{2}(\nabla{\bm{u}}+(\nabla{\bm{u}})^T)$ is the strain rate tensor. The heat conduction and diffusion flux are given as follows
\begin{equation}
\bm{q}_c=-\kappa \nabla {T},
\end{equation}
\begin{equation}
\bm{q}_d=-\sum_{l=1}^{2}\rho h_l(D_l \nabla{Y_l}-Y_l\sum_{m=1}^{2}D_m \nabla{Y_m}),
\end{equation}
where, $T$ is the static temperature of the fluids’ mixture. For
species $i=l$ or $m$, $h_i$ and $D_i$ are the individual enthalpy and the effective binary diffusion coefficient \cite{ramshaw1990self}, respectively. Additionally, the models for dynamic viscosity coefficient $\mu$, thermal conductivity $\kappa$ and the effective binary diffusion coefficient for the mixture are well documented by Tritschler \cite{Tritschler2014On} and Shanka \cite{shankar2014numerical}, and one can  also refer to our previous paper \cite{zeng2018turbulent} for details.

To close the governing equations, the equation of state (EOS) for the mixture of ideal gases is adopted in the present paper. Its formula is given by
\begin{equation}
p=\rho \overline{R} T,
\end{equation}
where $\overline{R}=\dfrac{R_u}{\overline{M}}$ is the gas constant of the mixture with $R_u$ being the universal gas constant and $\overline{M}=1/\underset{i=1}{\overset{2}{\sum}}\dfrac{Y_i}{M_i}$ being the mean molecular mass of the mixture. As proposed by Johnsen \cite{johnsen2012preventing}, the internal energy is related to pressure in the following form:
\begin{equation}
E=\dfrac{p}{(\gamma-1)\rho}+\dfrac{1}{2}\left | \bm{u}  \right |^2,
\end{equation}
where the specific heat ratio of mixture, $\gamma$, is given by
\begin{equation}
\dfrac{1}{\gamma-1}= \sum_{i=1}^{2}\dfrac{Y_i \overline{M}}{(\gamma_i-1)M_i} .
\end{equation}
In the above formulas, $M_i$ and $\gamma_i$ are, respectively, the molecular mass and specific heat ratio of species $i$ . 

\subsection{Numerical Method} 
In the framework of FV method, the semi-discretized form of Eq. (1) is used to update the cell-averaged physical states at the cell center, which is in the following form
\begin{equation}
\dfrac{d\overline{Q}}{dt}=-\dfrac{1}{\Omega}\sum_{nf=1}^{6}(\bm{F}_{c,nf}^*-\bm{F}_{v,nf}^*)\triangle{S}_{nf}+\overline{W}.
\end{equation}
In the above formula, $\overline{Q}$ is the average state of $Q$ at each cell center, $\bm{F}_{c,nf}^*=\bm{F}_c \cdot \bm{n}_{nf}$ and $\bm{F}_{v,nf}^*=\bm{F}_v \cdot \bm{n}_{nf}$ are, respectively, the numerical convective flux and viscous flux at the cell inteface $nf$ with area of $\triangle{S}_{nf}$ and unit outward normal vector of $\bm{n}_{nf}$, and $\overline{W}= \begin{bmatrix}0 &0 &0 &0 & \underset{nf=1}{\overset{6}{\sum}}\overline{\theta} (\bm{u}_{nf}\cdot{\bm{n}_{nf}} \triangle{S}_{nf}) \end{bmatrix}^T$ is the average source term of the cell with $\overline{\theta}$ being the corresponding average of the function of specific heat ratio for the mixture and $\bm{u}_{nf}$ being the fluids' velocity at cell interface.

Combined with the fourth-order MDCD reconstruction proposed by Wang \cite{wang2013low}, the Harten-Lax-van Leer-Contac (HLLC) Riemann solver for  quasi-conservative form of governing equations of multi-fluids, which is initially proposed by Abgrall \cite{abgrall1996prevent}, is used to calculate the numerical flux of convection, $\bm{F}_{c,nf}^*$. Additionally, Green’s theorem is used to integrate the numerical viscous flux, $\bm{F}_{v,nf}^*$. Moreover, to evaluate the average source term, the formula proposed by Johnse \cite{johnsen2012preventing} for the fluids' velocity at cell interface, $\bm{u}_{nf}$, is used, which is highly consistent with the HLLC Riemann solver for the numerical flux of convection. Once all the terms in Eq.(9) are evaluated, we update the flow states temporally  using the third-order total variation diminishing (TVD) Runge-Kutta method proposed by Shu \cite{jiang1996efficient}. In terms of the accuracy of the present simulation code, it has already been well demonstrated in our published work\cite{zeng2018numerical, zeng2018turbulent, wang2020consistent}. 
\section{Generating CCSW and Verification of CCSW/interface interaction} 
\subsection{Generating CCSW}
Generating cylindrically converging shock wave or spherically converging shock wave is somewhat complicated, no matter in experiment \cite{zhai2010generation, zhai2012parametric, eliasson2006focusing, dimotakis2006planar, vandenboomgaerde2011analytical, kjellander2012energy, liverts2016limiting} or in numerical simulation \cite{Lombardini2014Turbulent_a, bhagatwala2012interaction}. Based on Guderley's theory of converging shock wave \cite{guderley1942starke}, Lombardini and Pullin successfully set up the initial conditions for numerical study on the turbulent mixing driven by spherical implosions \cite{Lombardini2014Turbulent_a, Lombardini2014Turbulent_b}. However, there are some defects for Guderley's method in generating converging shock wave since the condition for the validity of Guderley's theory is that the shock wave must be strong enough. Another efficient method of generating converging shock wave is based on the theory of shock tube, which is initially used by Bhagatwala and Lele \cite{bhagatwala2012interaction} to generate the spherically converging shock wave. In this method, one just needs to increase the pressure and density ratios at a specific radius to generate the "pure" converging shock wave without a following contact surface. In this paper, we follow the approach of Bhagatwala and Lele \cite{bhagatwala2012interaction} to generate CCSW. 

According to the theory of shock tube, the region of low pressure gas (initially in static state with thermal states of pressure $p_1$, density $\rho_1$, and the ratio of specific heats $\gamma$) and the region of high pressure gas (initially in static state with thermal states of pressure $p_4$, density $\rho_4$, and the same ratio of specific heats $\gamma$) are initially separated by the diaphragm. As soon as the diaphragm is broken, a CCSW will propagate inward into the region of low pressure gas, while an expansion fan will propagate outward into the region of high pressure gas. To generate a "pure" CCSW with desired Mach number $M_{s0}$ and no following contact surface, the pressure $p_2$ and density $\rho_2$ of the gas in the intermediate region (the region between the CCSW and the expansion fan) at initial time should meet the following set of implicit formulations  \cite{bhagatwala2012interaction} 

\begin{widetext}
\begin{equation}
\dfrac{\rho_2}{\rho_1}=\dfrac{(\gamma+1)M_{s0}^2}{(\gamma-1)M_{s0}^2+2},
\end{equation}
\begin{equation}
\dfrac{p_2}{p_1}=1+\dfrac{2\gamma}{\gamma+1}  (M_{s0}^2-1),
\end{equation} 
\begin{equation}
\dfrac{\rho_4}{\rho_1}=\dfrac{\rho_2}{\rho_1}\left \{1-\dfrac{a_1}{a_4}(\gamma-1)(\dfrac{\rho_2}{\rho_1}-1)\sqrt{\dfrac{1}{2\dfrac{\rho_2}{\rho_1} \left [\gamma+1-(\gamma-1)\dfrac{\rho_2}{\rho_1} \right ]}} \right \}^{-\dfrac{2}{\gamma-1}},
\end{equation} 
\begin{equation}
\dfrac{p_4}{p_1}=\dfrac{p_2}{p_1}\left \{1-\dfrac{a_1}{a_4}(\gamma-1)(\dfrac{p_2}{p_1}-1)\sqrt{\dfrac{1}{2\gamma \left [2\gamma+(\gamma+1)(\dfrac{p_2}{p_1}-1) \right ]}}  \right \}^{-\dfrac{2\gamma}{\gamma-1}}.
\end{equation} 
\end{widetext}
In the above equations, $\phi_4/\phi_1$ and $\phi_2/\phi_1$, where $\phi \in \{\rho, p\}$, are jump ratios across the initial
diaphragm and the desired CCSW, respectively, and $a_i=\sqrt{\gamma p_i/\rho_i}$ (with $i=1,2$ or $4$) is the sound speed in the corresponding region.

In order to demonstrate the efficiency of the above method in generating CCSW, we set up a two-dimensional (2D) numerical simulation and verify its results using Guderley's theory \cite{guderley1942starke}. The initial radius of circular shock (corresponding to the diaphragm position) for this simulation is $R_0=$0.5 $m$. To generate a converging shock wave with initial Mach number of $M_{s0}=1.5$, the thermal states of initially static gas (air with specific heat ratio $\gamma=1.4$) inside the circular shock wave is set to be: $\rho_1=1.225\ kg/m^3$, $p_1=101325 \ pa$, $a_1=340.3\ m/s$. The corresponding thermal states of initially static air outside the circular shock wave, $\rho_4,\ p_4$ and $a_4$, can be given by iteratively solving Eq.(10)-Eq.(13). Once these initial conditions are given, a converging shock wave and a expansion fan will propagate inward and outward respectively from the initial position. The thermal states of the air in the intermediate region at the very beginning time, $\rho_2,\ p_2$ and $a_2$, can also be given by Eq.(10)-Eq.(13).

Obviously, the strength of the converging shock wave will be enhanced during its propagation since the area of shock surface is decreased. The Mach number ($M_s$) of the converging shock wave at any radius $r$ during its inward propagation in our numerical simulation can be derived from the following formulation \cite{anderson2003modern}
\begin{equation}
\dfrac{p_b}{p_1}=1+\dfrac{2\gamma}{\gamma+1}(M_s^2-1),
\end{equation}
where, $p_b$ is the pressure behind the inward propagating CCSW at the corresponding radius. 

Additionally, based on the theory of Guderley \cite{guderley1942starke}, during the inward propagation for a pure converging shock wave initially placed at the radius of $R_0$, its radius at given time $t$, $r(t)$, can be addressed as
\begin{equation}
r(t)=R_0(1-\dfrac{t}{t_0})^\alpha .
\end{equation}
In the above equation, $t_0$ is the total propagation time from the initial radius to the center for converging shock wave, which is about 0.865 $ms$ in the present simulation. Additionally, for pure CCSW propagating in gas with $\gamma=1.4$, the Guderley exponent $\alpha \approx 0.835$. Based on Eq.(15), one can directly derive the theoretical formula for Mach number of converging shock wave during its inward propagation \cite{si2015experimental}
\begin{equation}
M_s=\dfrac{1}{a_1}\dfrac{\alpha R_0}{t_0} \left(\dfrac{r}{R_0} \right)^{ \left(\alpha-1 \right)/\alpha},
\end{equation}
where, $a_1$ is the sound speed of gas inside the CCSW.
 
Based on the above formulations, both $r(t)$ and $M_s$ would be sensitive to the value of $\alpha$. Particularly, when the shock is propagating near to the center, a marginal change of $\alpha$ would result in a dramatical variation of $r(t)$ and $M_s$. In order to take the effect of $\alpha$ into account, we additionally chose another two values of Guderley exponent $\alpha^{\pm}=\alpha(1\pm 5\%)$, with $\alpha= 0.835$, for our comparation study. Fig.1 and Fig.2 compare the evolution of nondimensionalized radius $ln \left[{r(t)}/{R_0} \right]$ versus nondimensionalized time $ln \left(1- {t}/{t_0}\right)$ and the evolution of Mach number $M_s$ versus $r$ of CCSW in our numerical simulation with the corresponding results obtained from the theory of Guderley \cite{guderley1942starke}, respectively. As shown in Fig.1 and Fig.2, the evolutions of nondimensionalized radius and Mach number in our simulation agree well with those of Guderley's theory when the Guderley exponent is in marginal range of $[\alpha^-, \alpha^+]$, which could demonstrate that the method used by Bhagatwala and Lele \cite{bhagatwala2012interaction} to generate the spherically converging shock wave is also efficient in generating CCSW.

\begin{figure}
\centering
\includegraphics[width=0.9\linewidth]{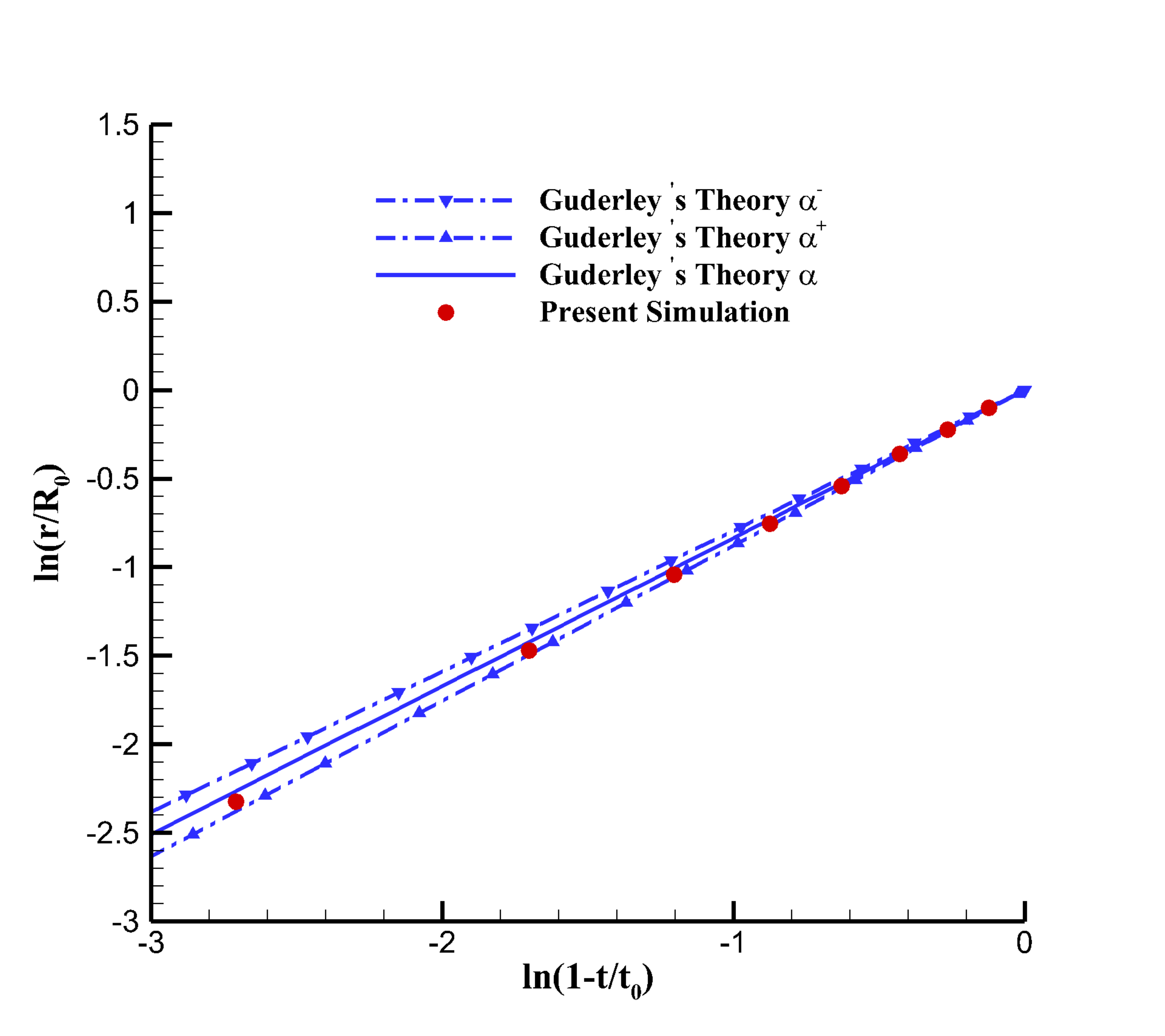}
\caption{\label{fig:fig1} Evolution of 
nondimensionalized radius $ln \left[{r(t)}/{R_0} \right]$ versus nondimensionalized time $ln \left(1- {t}/{t_0}\right)$.}
\end{figure}

\begin{figure}
\centering
\includegraphics[width=0.9\linewidth]{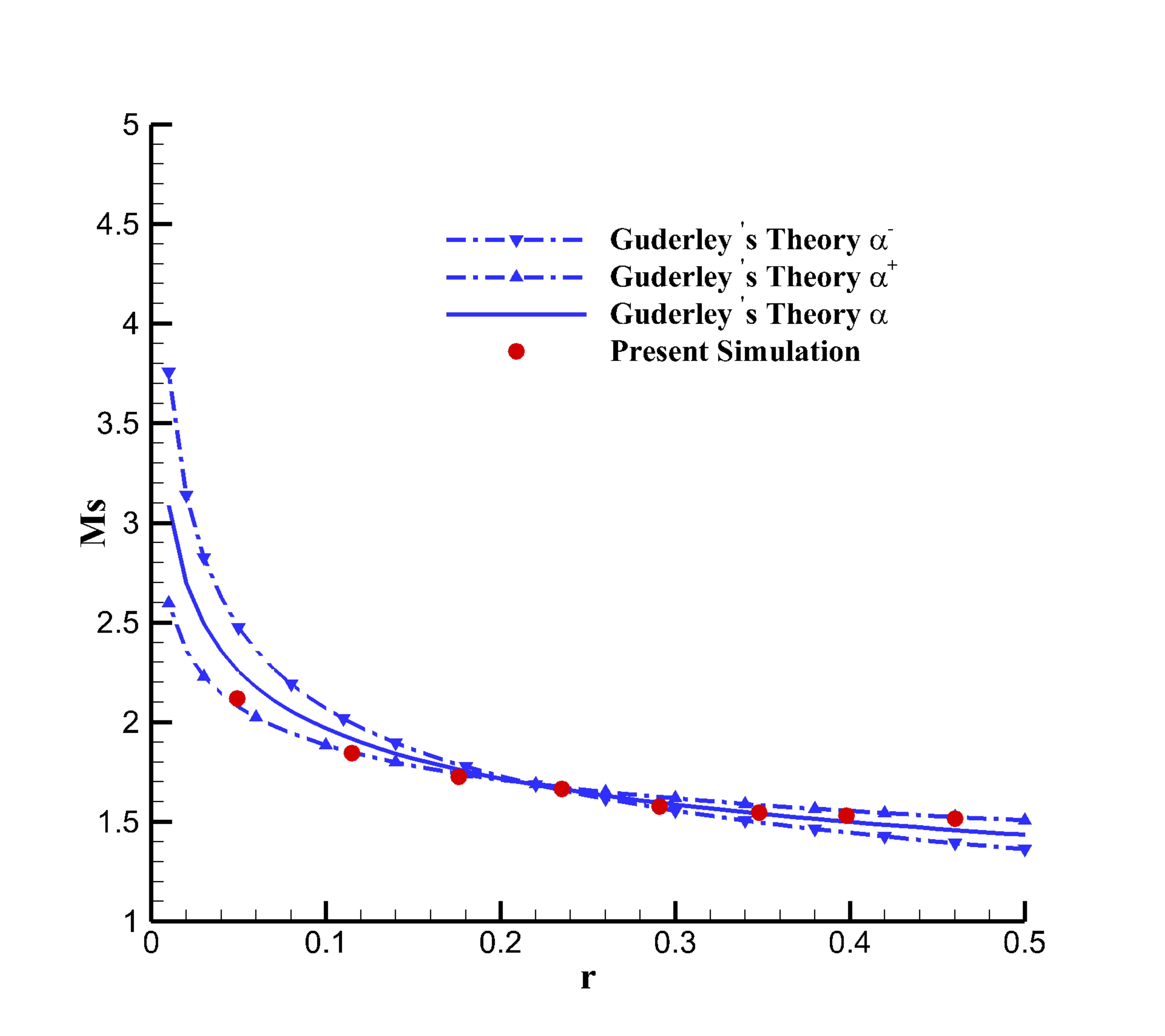}
\caption{\label{fig:fig2} Evolution of Mach number $M_s(r)$ versus $r$ .}
\end{figure}

\subsection{Verification of CCSW/interface interaction: Studies on the amplitude growth of CCSW driven interfaces}
One of most important features for RMI flow is the growth of perturbation amplitude. Due to the impulsive acceleration of incident shock wave, the initial perturbation will be stretched, resulting in the possible linear growth of perturbation amplitude at early stage. Initially studied by Richtmyer \cite{richtmyer1960taylor} and later confirmed experimentally by Meshkov \cite{meshkov1969instability}, the growth rate of perturbation amplitude for interface driven by planar shock wave is extensively studied during past three decades \cite{ZhangNonlinear_1997, VandenboomgaerdeNonlinear_2002, MatsuokaNonlinear_2003, SohnSimple_2003}. However, the perturbation growth of interface driven by converging shock wave remains to be an attractive topic with many open issues. Based on the CCSW generated by the method mentioned above, using 2D numerical simulations,  we try to explore some features of perturbation growth for CCSW driven interface before re-shock in this subsection. In addition to numerical simulations, the theoretical model proposed by Mikaelian \cite{mikaelian2005rayleigh} is also used to predict the amplitude growth of CCSW driven interface for further mutual confirmation.

In the present numerical studies of amplitude growth, the initial perturbation of gases interface is in the following form
\begin{equation}
\eta_0=r_0-a_0\left[ 1.0-cos(n\varphi) \right], 
\end{equation}
where, $r_0$ is the initial radius of  the outer interface (radius of crest at initial time), $a_0$ is the initial amplitude, $n$ is the azimuthal mode number, and $\varphi$ is the azimuthal angle. 
To make our results be more general, we take the effects of initial mode and initial amplitude into account. Therefore, three cases with different initial azimuthal mode number and/or initial amplitude are studied. The parameters of initial perturbations for the three cases are well listed in Table I. 
\begin{table}[H]
\caption{\label{tab:table1} Parameters of initial perturbations for study of amplitude growth}
\begin{ruledtabular}
\begin{tabular}{ccddd} Case Index &\mbox{$r_0\ (m)$} &\mbox{$a_0\ (cm)$}&\mbox{n}\\
\hline
I    &0.38 &1.0  &8  \\
II   &0.38 &2.0  &8  \\
III  &0.38 &1.0  &12 \\
\end{tabular}
\end{ruledtabular}
\end{table}

Moreover, the CCSW with desired Mach number of $M_{s0}=1.5$ is initially placed at radius of $R_0=0.4m$. The gas in the region between gases interface and the CCSW is air with initial thermal states of $\rho_1=1.225\ kg/m^3$, $p_1=101325 pa$, $a_1=340.3m/s$ and $\gamma_1=1.4$. Once more, the initial thermal states of gas (air) outside the CCSW is iteratively solved based on Eq.(10)-Eq.(13). Additionally, the gas inside the initial interface is a mixture of air ($20\%$ in mass fraction) and sulphur hexafluoride (SF$_6$, with specific heat ratio $\gamma_2=1.1$) under the condition that both temperature and pressure are in equilibrium states at the gases interfaces. The velocity for the gases in the whole computational domain is zero at the initial time. It would take $0.04\ ms$ for the CCSW to initially strike the gases interface when its Mach number approximately reaches 1.51. For boundary conditions, a viscous circular wall with radius of $1cm$ is placed around the center. Additionally, the circular boundary at the outside of the computational domain is large enough for the propagations of all possible waves during the durations of simulation. Consequently, a zero-gradient boundary condition is used for the outer boundary. A body-fitted mesh is used for all three simulations, with 4096 cells in azimuthal direction and 2440 cells in radial direction. The amplitude of interface in numerical simulations is then given by $a(t)\_\textsc{CFD}\approx \left[ r_{crest}(t)-r_{trough}(t)\right]/2$, where $r_{crest}(t)=max\{r_{Y_{SF_6}=0.5}\}$ and $r_{trough}(t)=min\{r_{Y_{SF_6}=0.5}\}$ are, respectively, the radius of the outer interface (crest) and inner interface (trough) of simulation results as shown in Fig.3.
\begin{figure}
\centering
\includegraphics[width=0.9\linewidth]{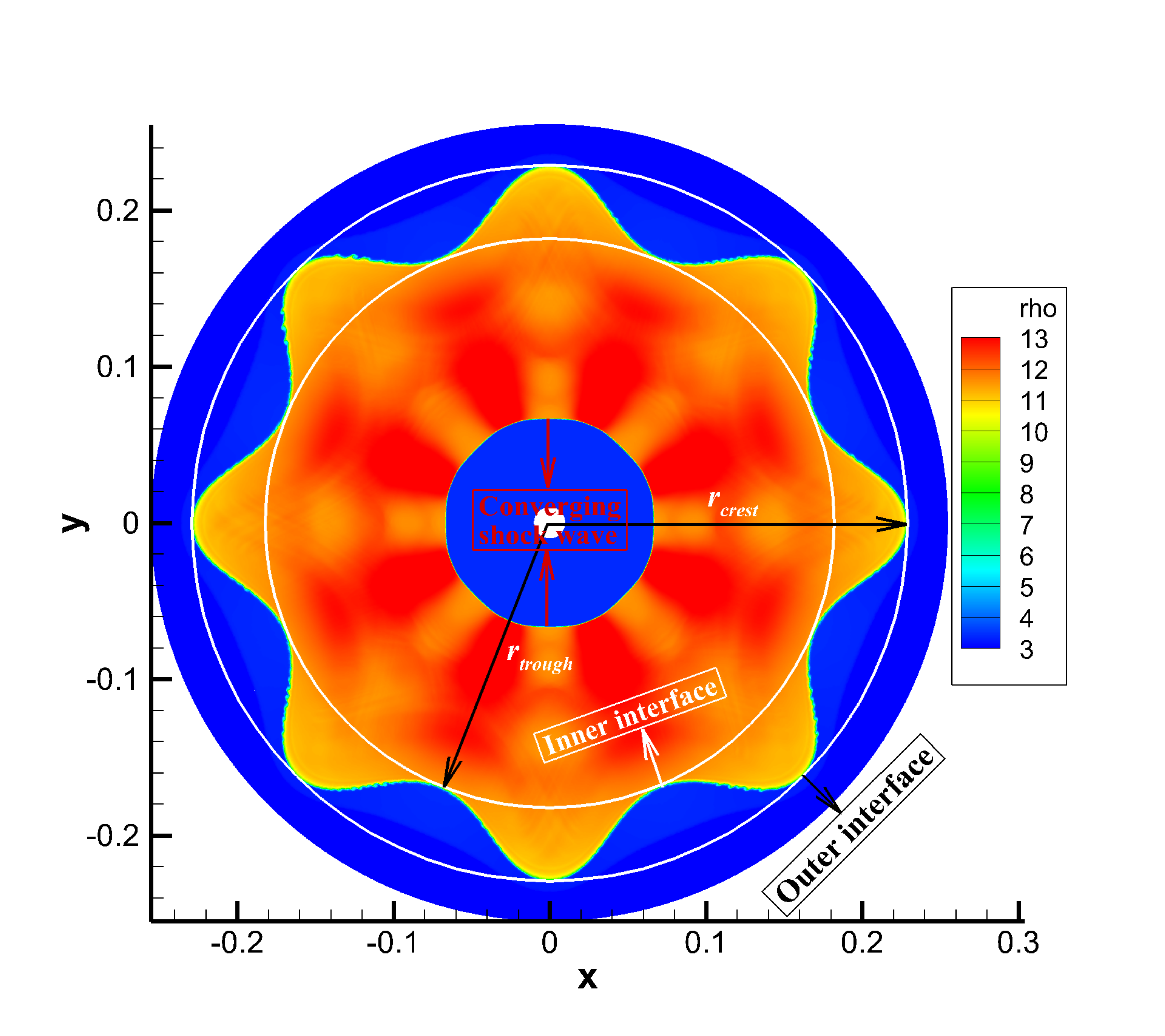}
\caption{\label{fig:fig3} Crest radius and trough radius in simulations.}
\end{figure}

As an alternative to numerical simulations, linear models are also widely used for predicting the perturbation development of CCSW driven gases interface \cite{mikaelian2005rayleigh, ding2017measurement, liu2019richtmyer, luo2019nonlinear, suponitsky2014richtmyer}. Based on the pioneered work of Bell \cite{bell1951taylor} and Plesset \cite{plesset1954stability}, Mikaelian modelled the amplitude growth rate of interface (with small ratio of initial amplitude to the initial wave length) driven by CCSW as follows \cite{mikaelian2005rayleigh}
\begin{equation}
\frac{d^2a(t)}{dt^2}=-2\frac{\dot{r}}{r}\frac{da(t)}{dt}+(nA-1)\frac{\ddot{r}}{r}a(t). 
\end{equation} 
In the above formula, $n$ is the mode number in azimuthal direction, $A=(\rho_{in}-\rho_{out})/(\rho_{in}+\rho_{out})$ is the Atwood number with $\rho_{in}$ and $\rho_{out}$ respectively being the density inside and outside the already shocked interface. Additionally, $a(t)$ and $r$ are, respectively, the amplitude and average radius of the gases interface at time $t$. There is no analytical formula for $r(t)$, consequently, it is approximately given by $r(t)=\left[ r_{crest}(t)+r_{trough}(t)\right]/2$ under the assumption that the perturbation amplitude of interface is small. Obviously, for $t>t_0^+$ (where $t_0^+$ is the time when the incident CCSW passes through the trough of interface), Eq.(18) can be integrated in the following form \cite{ding2017measurement, liu2019richtmyer, luo2019nonlinear}
\begin{equation}
a(t)-a_0^+=
\dot{a}_0^+{\overline{r}_0}^2\int_{t_0^+}^{t}\dfrac{1}{r^2}dt+(nA-1)\int_{t_0^+}^{t}\dfrac{\int_{t_0^+}^{t}a(\tau)r\ddot{r}d\tau}{r^2} dt,
\end{equation} 
where $a_0^+\approx a_0(1-V_{si}/V_{is})$ and $\dot{a}_0^+=a_0^+(nA-1)V_{si}/\overline{r}_0$ are, respectively, the initial amplitude and growth rate of amplitude at time $t_0^+$, and $\overline{r}_0 \equiv r_0-a_0 \approx r_0$ for small $a_0$. Additionally, in the above formulations, $V_{is}$ is the velocity of CCSW when it initially strikes the gases interface and $V_{si}$ is the velocity jump of the shocked gases interface. There are two terms on the right hand side of Eq.(19), of which the first term denotes the Bell-Plesset (BP) effect on the growth of amplitude and the second term denotes Rayleigh-Taylor (RT) effect on the growth of amplitude \cite{ding2017measurement, liu2019richtmyer, luo2019nonlinear}. These two terms are corresponding to the effects of the first and second term on the right hand side of Eq.(18), respectively.

For the given $a_0^+$ and $\dot{a}_0$ based on initial conditions, the perturbation amplitude of Mikaelian's model can be obtained by directly integrating Eq.(18) using standard 4$^{th}$ order Runge-Kutta method. In order to identify the BP effect and RT effect on the growth of perturbation more clearly, we integrate Eq.(18) in two ways. In what follows, $a(t)\_$BP stands for the theoretical amplitude calculated by only integrating the first term of Eq.(18) on the right hand side, while $a(t)\_$BPRT stands for the theoretical amplitude calculated by integrating both terms of Eq.(18) on the right hand side.

Obviously, the BP effect is fully resulted from the geometric convergence effects on flows, since this term will be zero for the planar shock driven interface \cite{mikaelian2005rayleigh}. Additionally, as we will discuss below, Rayleigh-Taylor effect will not be enhanced by the geometric convergence at early stage. However, it, indeed, plays an important role in the growth of perturbation amplitude at later stage. Fig.4 shows the evolutions of perturbation amplitude obtained from numerical simulations as well as theoretical model of Mikaelian for all three cases. As shown in Fig.4, the results  of Mikaelian's theoretical model agree well with our numerical results at early stage for all three cases (see Region I for each case). Additionally, there are slight differences between $a(t)\_$BP and $a(t)\_$BPRT during this stage, which means that BP effect is the dominant factor for the growth of perturbation amplitude at early stage. However, at later stage (see Region II for each case), the evolutions of $a(t)\_$BP will increasingly diverge from the numerical results, while the evolutions of $a(t)\_$BPRT can still mimic the numerical results although there are some differences between them. These results indicate that the RT effect plays an important role in in the development of perturbation at later stage before re-shock (even overwhelms the BP effect at the very later stage before re-shock since the amplitudes for all three cases are decreasing at the end of Region II).
\begin{figure*}
\centering

\subfigure[Case I]{
\begin{minipage}[t]{0.5\linewidth}
\centering
\includegraphics[width=1.0\linewidth]{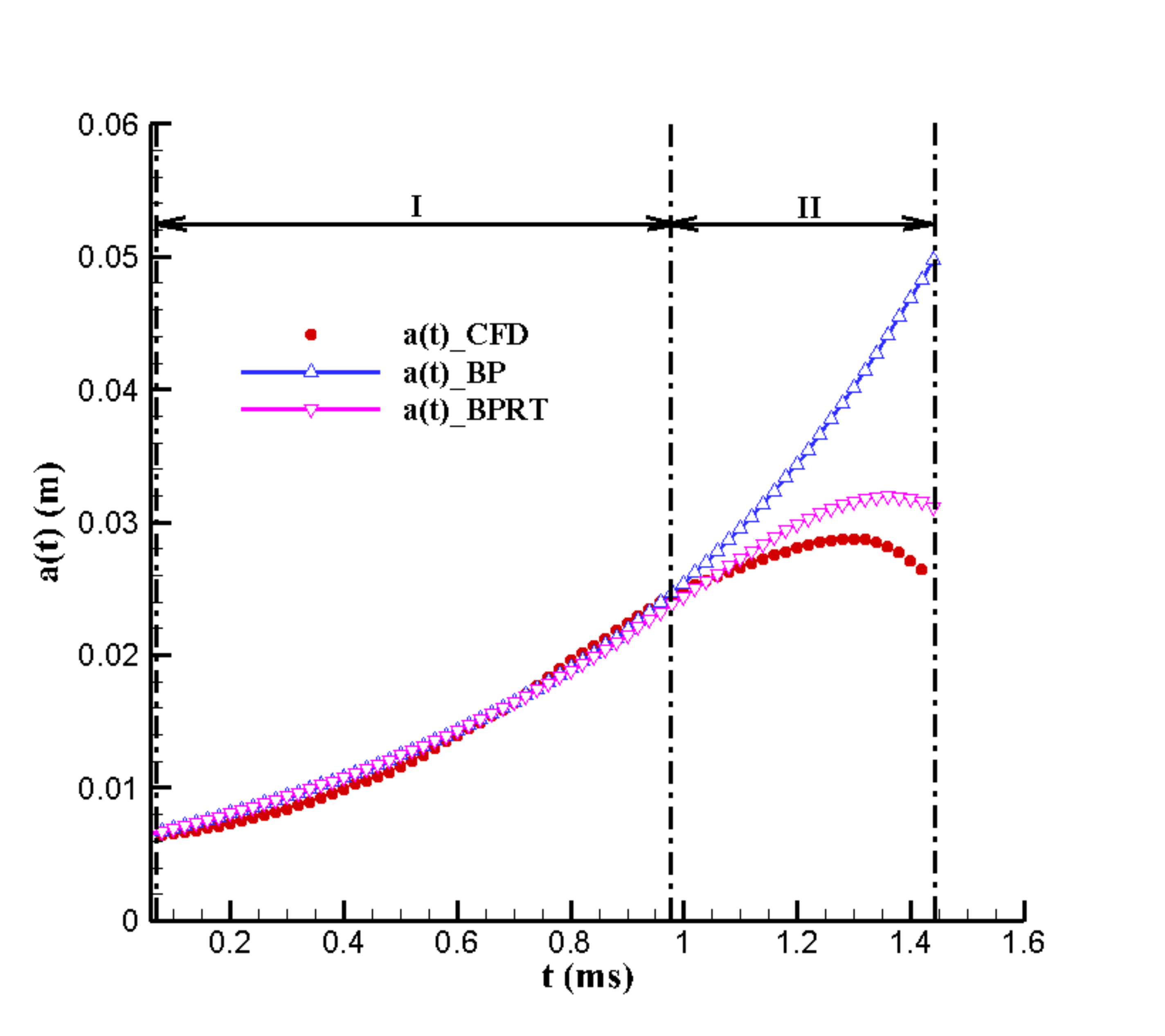}
\end{minipage}%
}%
\subfigure[Case II]{
\begin{minipage}[t]{0.5\linewidth}
\centering
\includegraphics[width=1.0\linewidth]{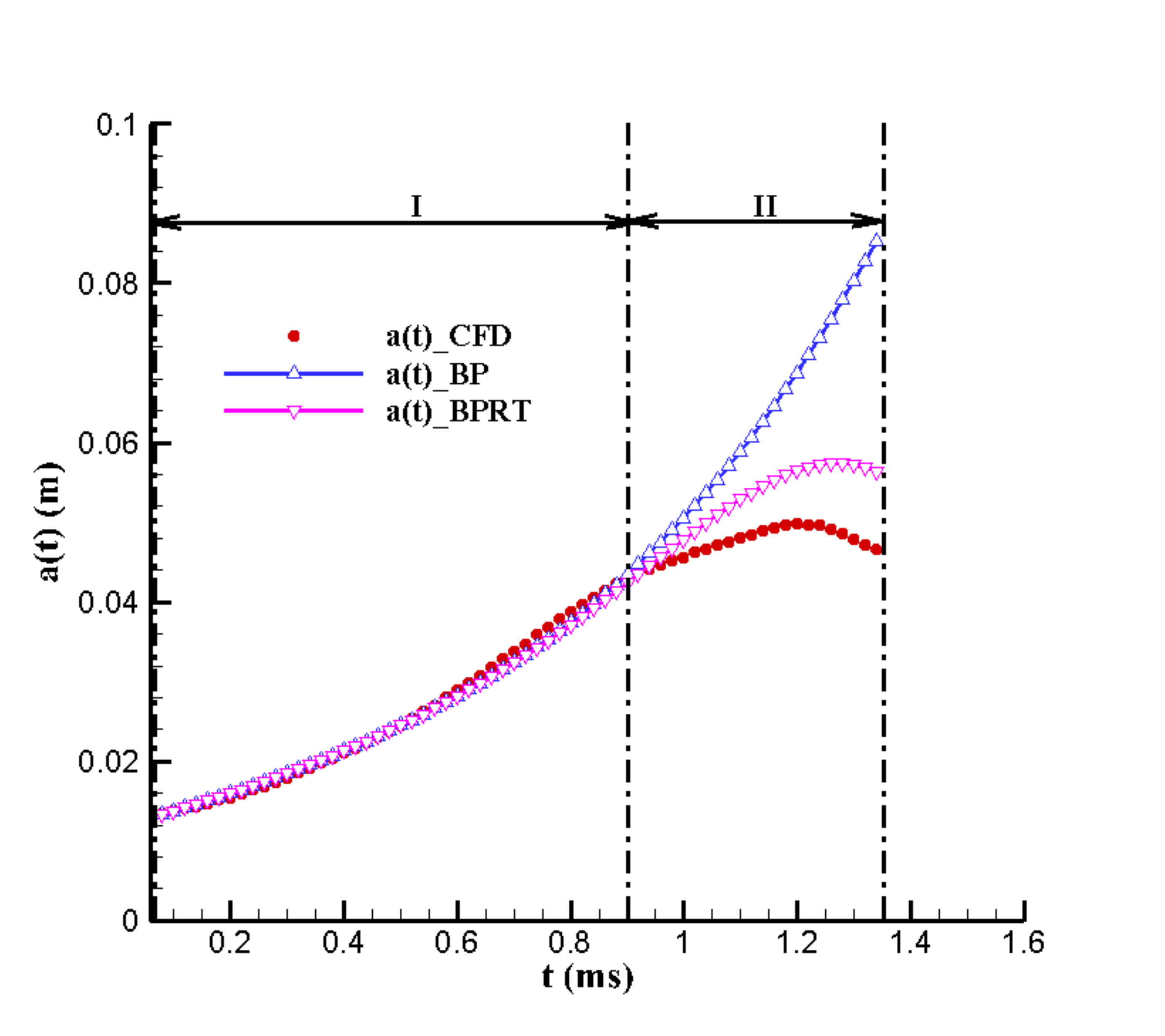}
\end{minipage}%
}%

\subfigure[Case III]{
\begin{minipage}[t]{0.5\linewidth}
\centering
\includegraphics[width=1.0\linewidth]{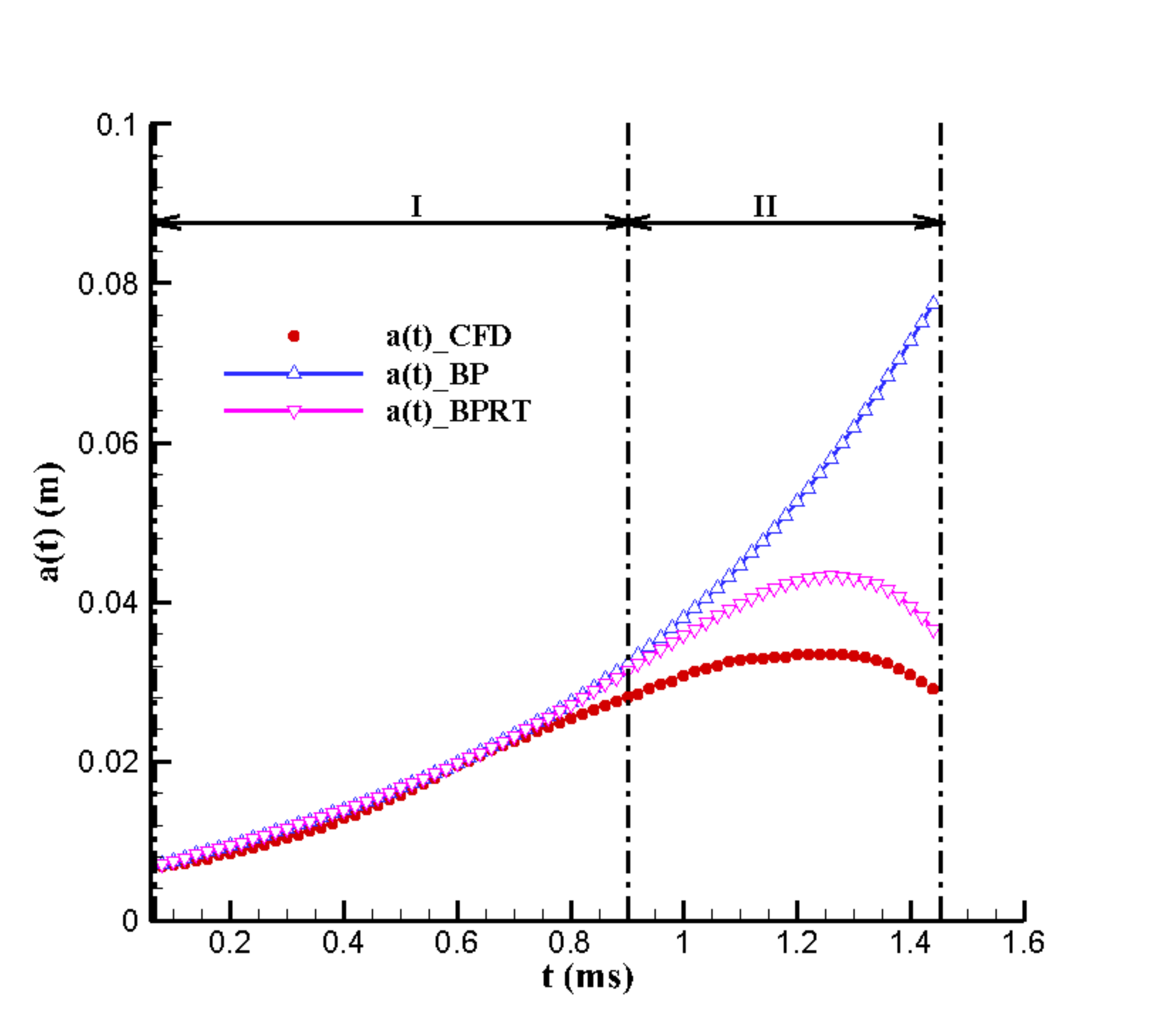}
\end{minipage}%
}%

\caption{\label{fig:fig4} Comparisons of amplitude growth between numerical results and theoretical results}
\end{figure*}

Moreover, according to the theoretical model of Mikaelian [see Eq.(19)], at the early stage when the BP effect dominates the amplitude growth, the nondimensionalized amplitude $a(t)/a_0$ can be approximately given by 
\begin{equation}
\dfrac{a(t)}{a_0} \approx (1-\dfrac{V_{si}}{V_{is}}) [1+C_1(nA-1) V_{si} \overline{r}_0].
\end{equation}
In the above formula, $\overline{r}_0$ and $C_1=\int_{t_0^+}^{t}\dfrac{1}{r^2}dt $ would become approximately constant since the initial amplitude is small. For all three cases in the present study, $V_{si}$ and $V_{is}$ are the same since the Mach number of the incident CCSW and the initial densities inside and outside the interface are the same. Consequently, during this stage, $a(t)/a_0$ should be approximately proportional to the wave number of initial interface. Fig.5 compares the evolutions of nondimensionalized amplitude for three numerical simulations (versus nondimensionalized time $t/t_{wall}$, where $t_{wall}=1.06\ ms$ is the approximate time when the CCSW strikes the inner wall boundary). As shown in Fig.5, due to the same wave numbers of initial perturbations, the nondimensionalized amplitudes of Case I and Case II  grow almost in the same way at the early stage. On the other hand, due to a larger wave number of initial perturbation, the nondimensionalized amplitude of Case III is larger than those of Case I and Case II during the corresponding stage. 
\begin{figure}
\centering
\includegraphics[width=1.0\linewidth]{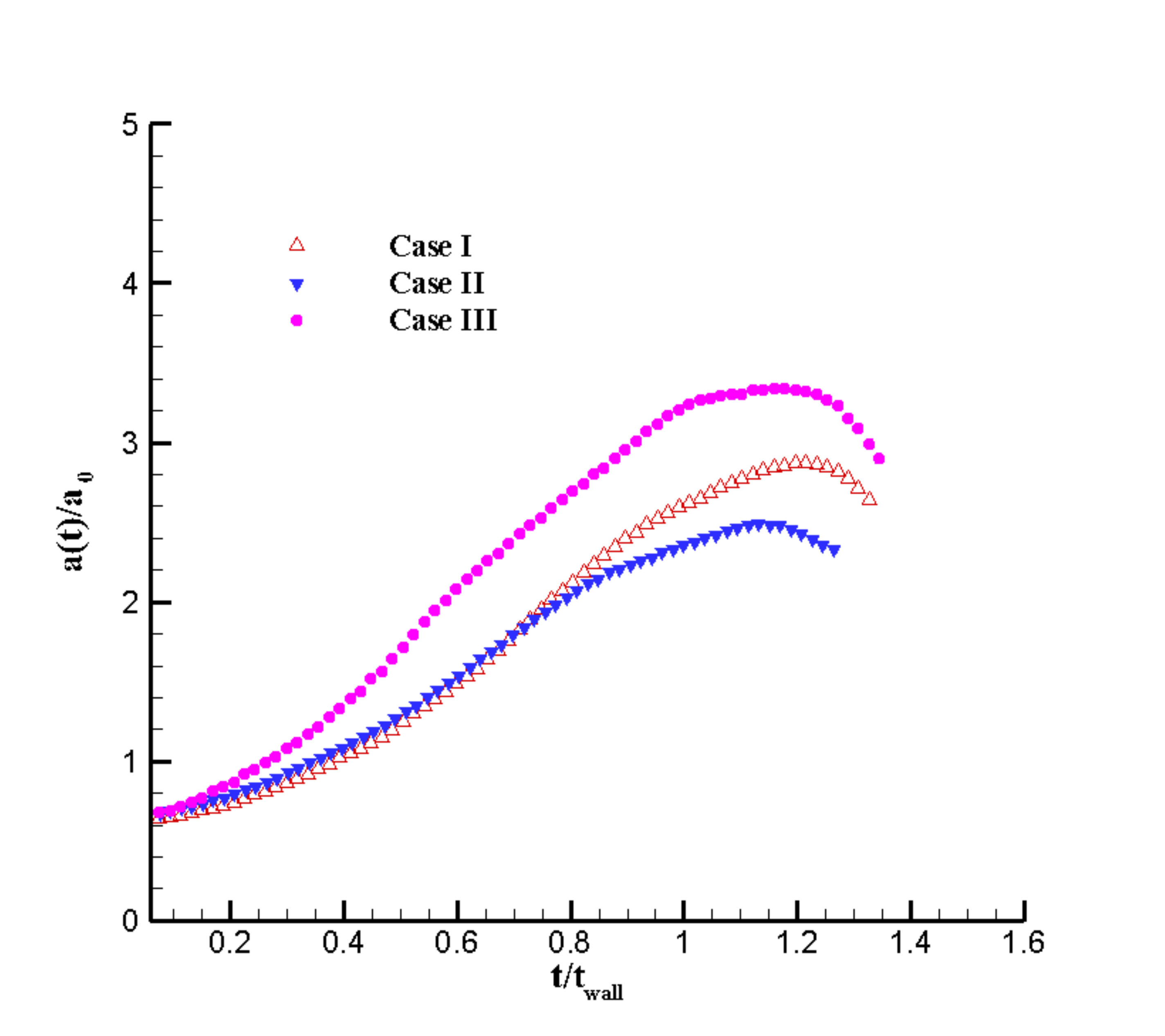}
\caption{\label{fig:fig5}  Evolutions of nondimensionalized amplitude for all three cases in numerical simulations.}
\end{figure}

According to the above analyses, we can see that, at early stage, the BP effect is the dominant factor of the  perturbation growth for CCSW driven interfaces, and the amplitude growth obtained from the theoretical prediction of Mikaelian's model agrees well with our numerical simulation results. However, at later stage, the RT effect becomes important and would even overwhelm the BP effect. Moreover, during the later stage, although there are some differences between the results of theoretical prediction (including RT effect) and our numerical simulations, the evolutions of $a(t)\_$BPRT can mimic the trend of our numerical results on the whole. In fact, such differences of amplitude growth have also been observed between the results of Mikaelian's theoretical model and shock tube experiments \cite{ding2017measurement, luo2019nonlinear}. There are several factors which can account for the above differences between the evolutions of $a(t)\_$BPRT and the results of numerical simulations/experiments. The main one is that the amplitude of perturbation becomes larger at later stage as the flows evolve. Consequently, the assumption of Mikaelian's model, which requires that the amplitude of perturbation should be small, is not quite valid anymore. Another one is that, in practice, we can not have the analytical formula for $r(t)$.  Consequently, the approximate formulation of $r(t)$ could introduce errors,  especially for the term of RT effect since it involves the second-order derivative of $r(t)$. In summary, all the results mentioned above, to some extent, not only could demonstrate the features of amplitude growth of perturbation for CCSW driven interfaces, but also can manifest that the CCSW derived from the aforementioned theory of shock tube indeed can be used for the studies of CCSW induced RMI flows.
 
\section{FLUIDS' MIXING OF CCSW DRIVEN INTERFACES}
The fluids' mixing in the mixing zone of gases interface is another crucial topic for shock-driven inhomogeneous flows. Better understandings of the mixing behaviors of shock-driven flows can shed light on the mechanisms of turbulent mixing \cite{zeng2018turbulent} as well as the turbulence modeling for such kind of flows \cite{livescu2009high, cabot2013statistical}. Recently, the mixing behaviors of planar shock driven interface are widely studied \cite{Hill2006Large, Thornber2010The, Tritschler2014On, Lombardini2012Transition, Thornber2017Late, Olson2014Comparison}. However, the corresponding behaviors of gases interface driven by converging shock wave, which are more important for some scientific disciplines such as ICF and supernova explosions, remain to be further investigated. In this section, we follow the approach of implicit large eddy simulations (ILES) to have a primary study on the mixing behaviors of three-dimensional (3D) gases interfaces driven by CCSW. Moreover, to take the effects of mode of initial perturbation on the mixing behaviors into account and to make our results be more general, two cases with different amplitudes and mode numbers are studied.

\subsection{Problems setup}
As remarked by Mikaelian \cite{mikaelian2005rayleigh}, the fluids' mixing of gases interface driven by converging shock wave would not happen dramatically if the wave length of initial perturbation is much larger than the amplitude of initial perturbation. In order to enhance the fluids' mixing at later stage after the gases interfaces are re-shocked, the initial perturbation for both cases in the present study are set to be a linear combination of "egg-carton" \cite{Hill2006Large}, which, in the cylindrical coordinate system, can be addressed as
\begin{equation}
\eta(\varphi,\ z)=\dfrac{r_0-a_0\times \left | cos(n_\varphi \varphi) cos(n_z z) \right | +\delta -r}{2\delta}.
\end{equation}
In the above formula, $r_0$ and $a_0$ are, respectively, the crest radius and the amplitude of the initial interface; $n_\varphi$ and $n_z$ are, respectively, the azimuthal mode number and axial mode number; $r=\sqrt{x^2+y^2}$ is the radius away from the center and $\varphi$ is the azimuthal angle. Combined with an initial diffusion layer (with thickness $\delta=0.01cm$) of the form proposed by Latini \cite{latini2007effects}, the mass fraction of SF$_6$ inside the CCSW can be addressed as
\begin{equation}
Y_{\text{SF}_6}=\left\{\begin{matrix}
 Y_{\text{SF}_{6},0}&  if\ \eta \geqslant 1.0& \\ 
Y_{\text{SF}_{6},0}\times(1.0-e^{\left | \eta \right |^8 ln \beta})&  if\  0 \leqslant \eta <1.0 & \\
 0&  if \eta<0&
\end{matrix},\right.
\end{equation}
where $\beta$ is the machine zero. According to the above formulation, at initial time, the mass fraction of $\text{SF}_6$ for the mixture of $\text{SF}_6$ and air inside the interface (the diffusion layer) is $Y_{\text{SF}_{6},0}$ ($Y_{\text{SF}_{6},0}=0.75$ for both cases), while the gas is pure air outside the interface. The parameters of the initial perturbations for both cases are listed in Table II. Furthermore, we initially place the CCSW with desired Mach number of $M_{s0}=1.5$ at radius of $R_0=0.4\ m$ for both cases. The thermal states of air in the region between the initial gases interface (diffusion layer) and the CCSW are the same as that we set in the second part of Section III, which will result in the same thermal states of air outside the initial CCSW. Moreover, for the present 3D simulations, we extend the axial width, $L_z$, to be 0.128 $m$. The configurations of initial flow field are shown in Fig.6. As for boundary conditions, a viscous cylindrical wall with radius of $1\ cm$ is placed around the center, and the periodic boundary condition is used along axial direction.  Additionally, in order to avoid the effects of waves reflected by the outer boundary, the cylindrical outer boundary is far away from the flow structure evolving region (the region with fine grid). To reduce computational costs, a hyperbolic mesh stretching is applied between the fine-grid domain and the outer boundary along the radial direction. The whole computational domain is discretized by a body-fitted mesh with total cells of $1024 \times 640 \times 128$ (azimuthal cells $\times$ radial cells $\times$ axial cells). 
\begin{table}
\caption{\label{tab:table2} Parameters of initial perturbations for CCSW induced interfacial fluids' mixing}
\begin{ruledtabular}
\begin{tabular}{ccddd} Case Index &\mbox{$r_0\ (m)$} &\mbox{$a_0\ (mm)$} &\mbox{$n_\varphi$}  &\mbox{$n_z$} \\ 
\hline
1   &0.38 &5.0  &32  &8 \\
2   &0.38 &10.0 &16  &4  \\
\end{tabular}
\end{ruledtabular}
\end{table}

\begin{figure}
\centering

\subfigure[Case 1]{
\begin{minipage}[t]{1.0\linewidth}
\centering
\includegraphics[width=1.0\linewidth]{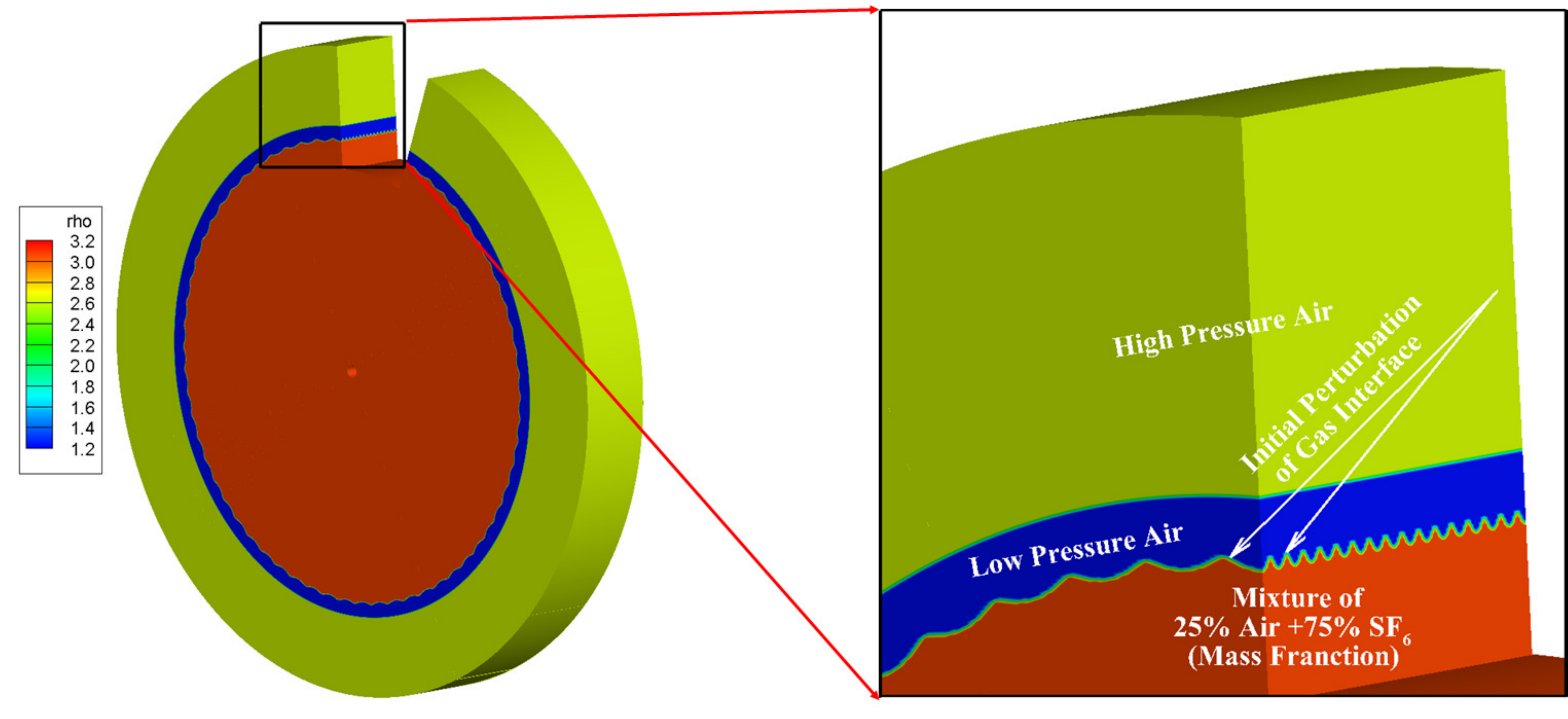}
\end{minipage}%
}%

\subfigure[Case 2]{
\begin{minipage}[t]{1.0\linewidth}
\centering
\includegraphics[width=1.0\linewidth]{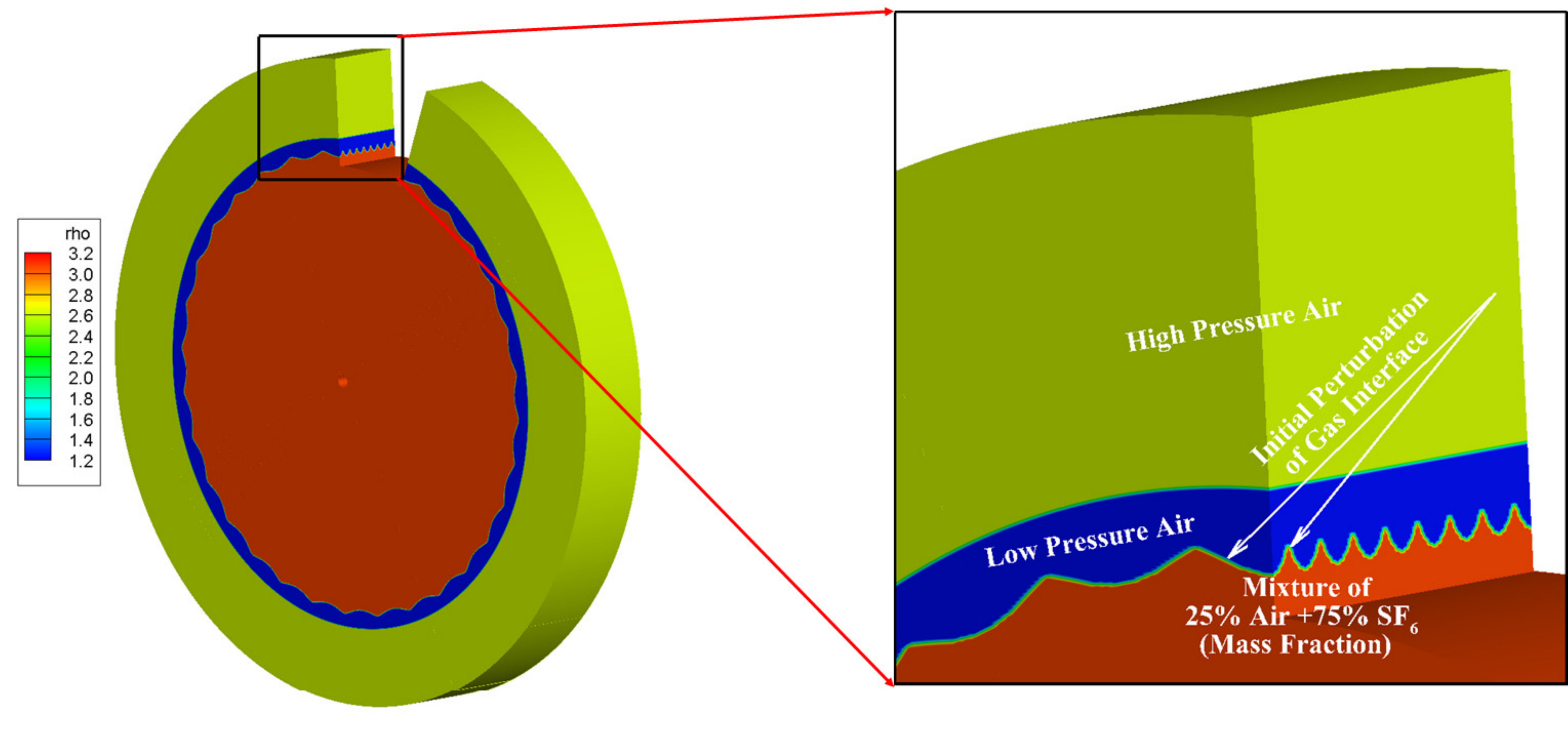}
\end{minipage}%
}%

\caption{\label{fig:fig6} Configurations of initial flow field for both cases.}
\end{figure}

Due to the moderate grid resolution, we hereby remark that the numerical dissipation would overweigh the physical one for the present simulations. Consequently, the present studies can be categorized as a class of ILES, in which the equations are implicitly filtered by the discretization and the numerical dissipation is treated as a surrogate for an explicit subgrid-scale model \cite{schilling2010high}. As remarked by Grinstein \cite{grinstein2013implicit} and Attal \cite{attal2015numerical}, although ILES could only resolve the length scales of turbulent mixing driven by advection and convective stirring, many studies show that they are indeed suited to (moderate) high-Reynolds-number flows in which shocks and interfaces are present \cite{Thornber2010The, gowardhan2011numerical, zhou2014estimating, thornber2011growth, thornber2012physics}. Moreover, our previous study on reshocked heavy gas curtain shows that the fluids' mixing behaviors of coarse grid (corresponding to the results of ILES) agree well with those of fine grid (corresponding to the results of direct numerical simulation) statistically \cite{zeng2018turbulent}, which, to some extent, manifests that our numerical method is appropriate for ILES.
\subsection{Results and Discussions}
\subsubsection{Wave patterns and flow structures evolutions}
The wave patterns of RMI induced by converging shock are more complicated than those of RMI induced by planar shock, which would affect the resulting evolutions of flow field to some extent. Fig.7 shows the typical wave patterns of CCSW induced RMI (results of Case 1 in the axial view). At early stage ($t=0.240\ ms$) shown in Fig.7(a), the reflected expansion fan (REF), which is associated with the incident CCSW and generated at the initial time, would propagate outward all the time. Additionally, as the results of interaction between the incident CCSW and initial interface, the reflected shock wave (RSW) will always propagate outward while the first transmitted shock wave (FTSW) will propagate inward initially. Obviously, the REF and the RSW would not impose much effects on the later evolutions of the flows since they propagate outward all the time and would not be reflected by the outer boundary which is large enough. However, the FTSW will be reflected by the inner wall boundary and re-impact the gases interface, which will result in the second transmitted shock wave (STSW) and the reflected rarefaction wave (RRW) at later time ($t=1.740\ ms$) as shown in Fig.7(b).

To identify the propagations of shock waves and their effects on the evolutions of flow structures more clearly, Fig.8 and Fig.9 show the details of flow evolutions and the propagations of shock waves for Case 1 and Case 2, respectively. As shown in Fig.8(a) and Fig.9(a), at early stage ($t=0.840 ms$), the gases interfaces will move inward with growth of perturbation amplitudes since the FTSWs will induce inward radial velocities. However, the inward propagating FTSWs will be reflected by the inner wall boundary. Then, the reflected FTSWs will propagate outward and begin to re-impact the gases interfaces, which are well shown in Fig.8(b) and Fig.9(b). After the reflected FTSWs re-impact the gases interfaces, the gases interfaces will move outward since the resulting STSWs will propagate outward and induce outward main radial velocities [see Fig.8(c)-(d) and Fig.9(c)-(d)]. Moreover, based on the evolutions shown in Fig.8 and Fig.9, we can see that the flow structures are characterized by the growth of perturbation amplitudes and the fluids' mixing is not intensive at early stage (Therefore, we just show one quarter of the flow fields). However, the fluids' mixing is dramatically enhanced by the second RMI after the gases interfaces are re-shocked [see the SF$_6$ mass fraction iso-surface shown in Fig.8(c)-(d) and Fig.9(c)-(d)].  

The morphological patterns of waves and the motions of flow structures (the evolutions of positions of inner and outer interfaces) mentioned above for both cases are depicted quantitatively in Fig.10. As shown in Fig.10, there are some unique features for the CCSW induced RMI flows. One is that the FTSW will move faster as it propagates inward due to the deformation (decreasing area) of shock surface. This is quite different from the RMI flows induced by planar shock wave, since, for planar shock driven RMI flows, the transmitted shock wave will propagate forward with a nearly constant velocity \cite{latini2007high, orlicz2009mach}. Another feature is that the movements of inner and outer gases interfaces are nonlinear versus time before re-shock, while, for planar shock driven RMI flows, the shocked interface will move forward with an approximately constant velocity as well \cite{latini2007high, orlicz2009mach, balakumar2008simultaneous}. Obviously, the nonlinear movements of inner and outer gases interfaces before re-shock will result in the nonlinear growth of perturbation amplitudes or mixing zone width at the very beginning (see Fig.11 shown below). Moreover, we hereby remark that the model of Mikaelian mentioned in the second part of Sections III would not be applicable to the nonlinear growth of perturbation amplitudes for these two cases, since, for both cases, their ratios of initial amplitude to initial wave length are not small enough. 

\begin{figure}
\centering

\subfigure[$t=$0.240\ $ms$]{
\begin{minipage}[t]{0.72\linewidth}
\centering
\includegraphics[width=1.0\linewidth]{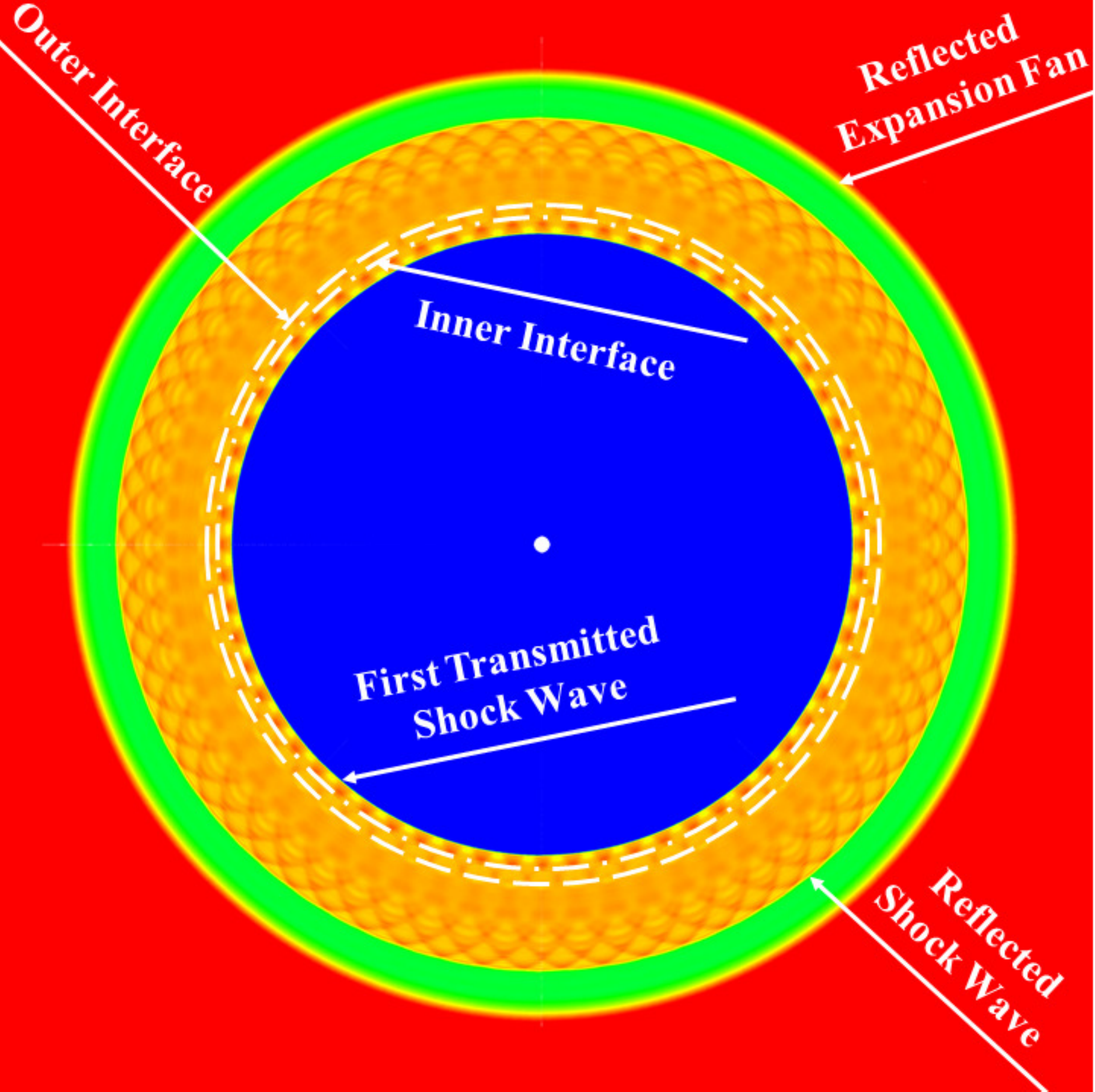}
\end{minipage}%
}%

\subfigure[$t=$1.740\ $ms$]{
\begin{minipage}[t]{0.72\linewidth}
\centering
\includegraphics[width=1.0\linewidth]{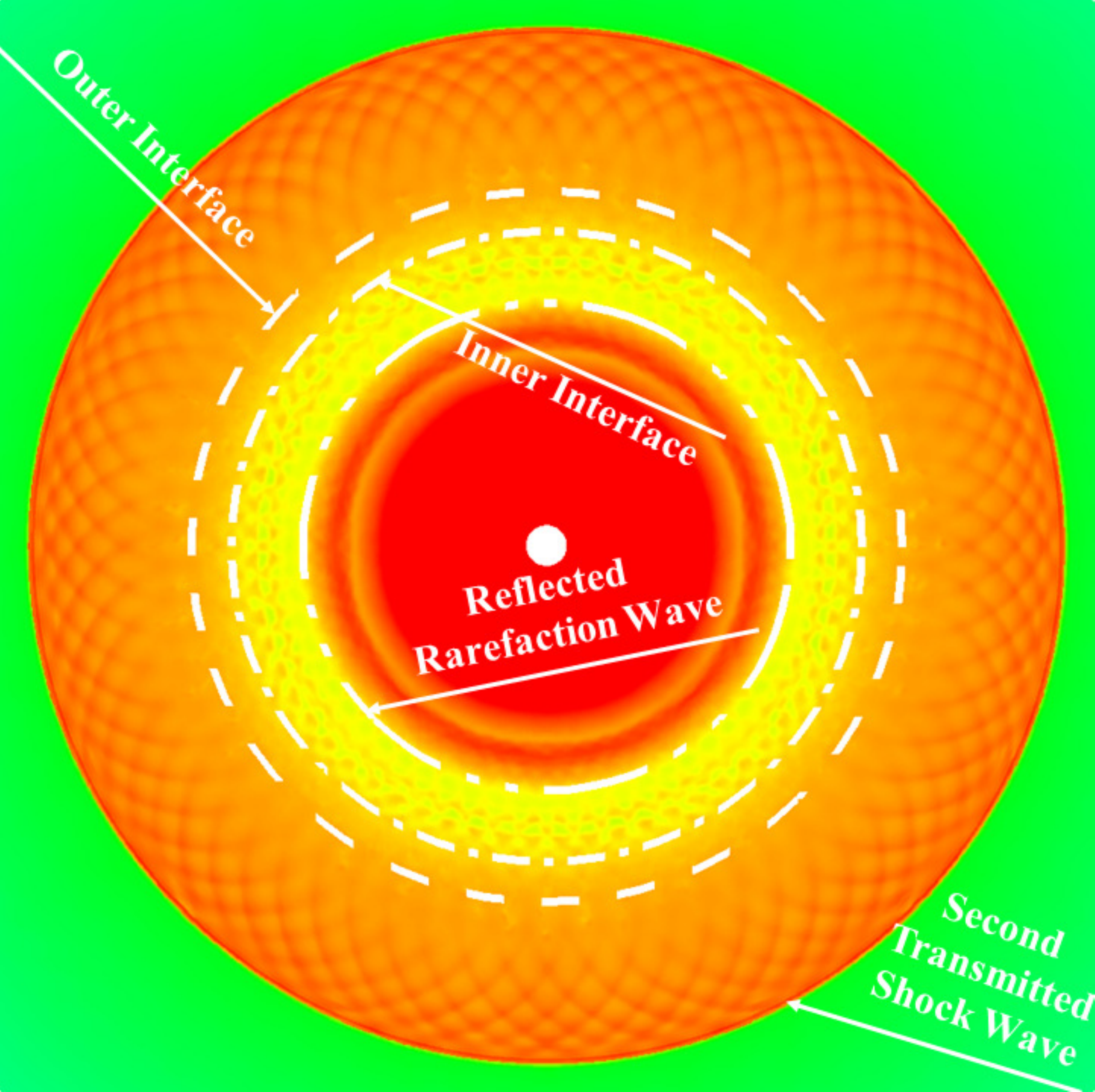}
\end{minipage}%
}%

\caption{\label{fig:fig7} Typical wave patterns of the CCSW induced RMI.}
\end{figure}

\begin{figure*}
\centering

\subfigure[$t=$0.840\ $ms$]{
\begin{minipage}[t]{0.5\linewidth}
\centering
\includegraphics[width=0.85\linewidth]{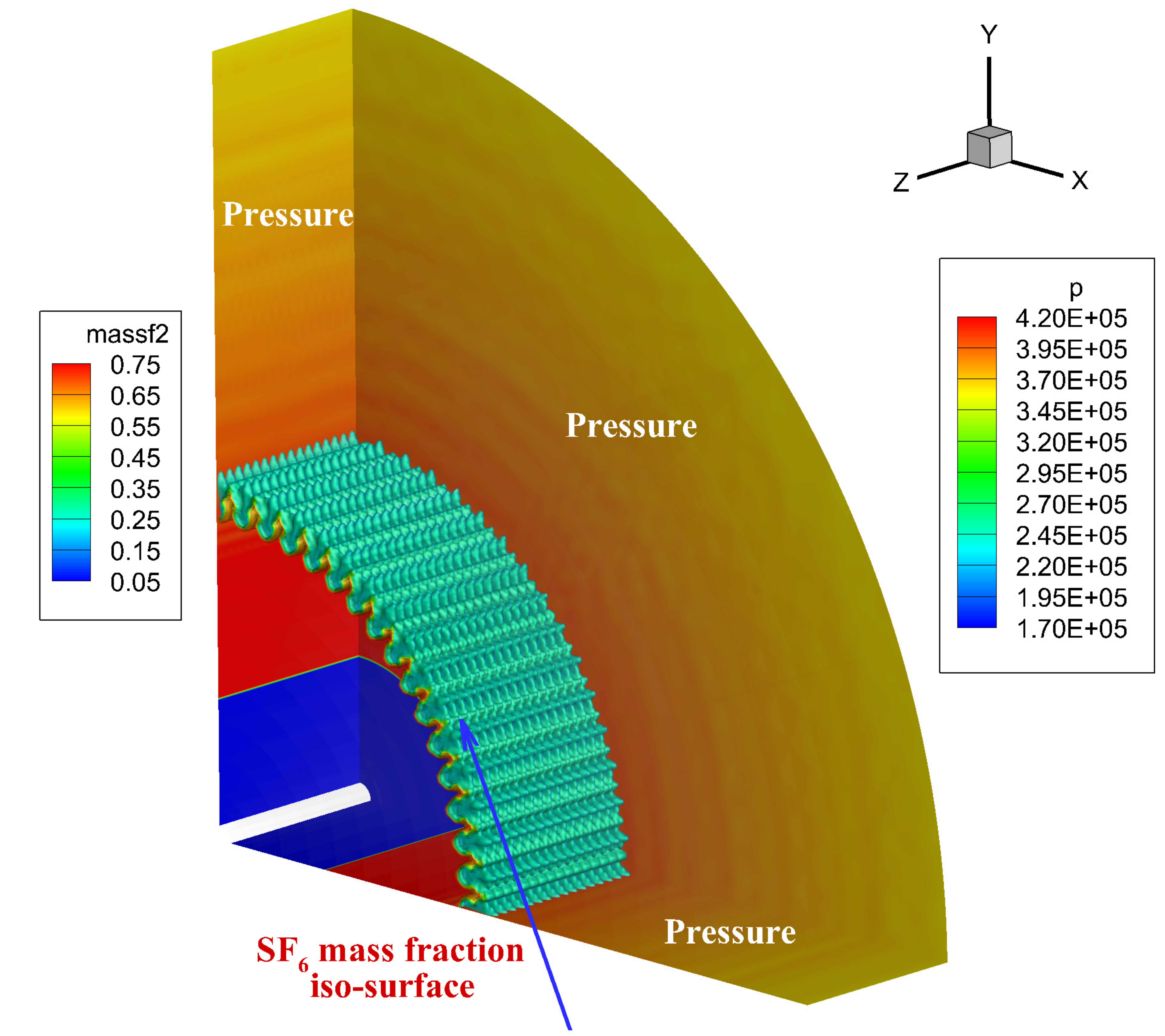}
\end{minipage}%
}%
\subfigure[$t=$1.440\ $ms$]{
\begin{minipage}[t]{0.5\linewidth}
\centering
\includegraphics[width=0.85\linewidth]{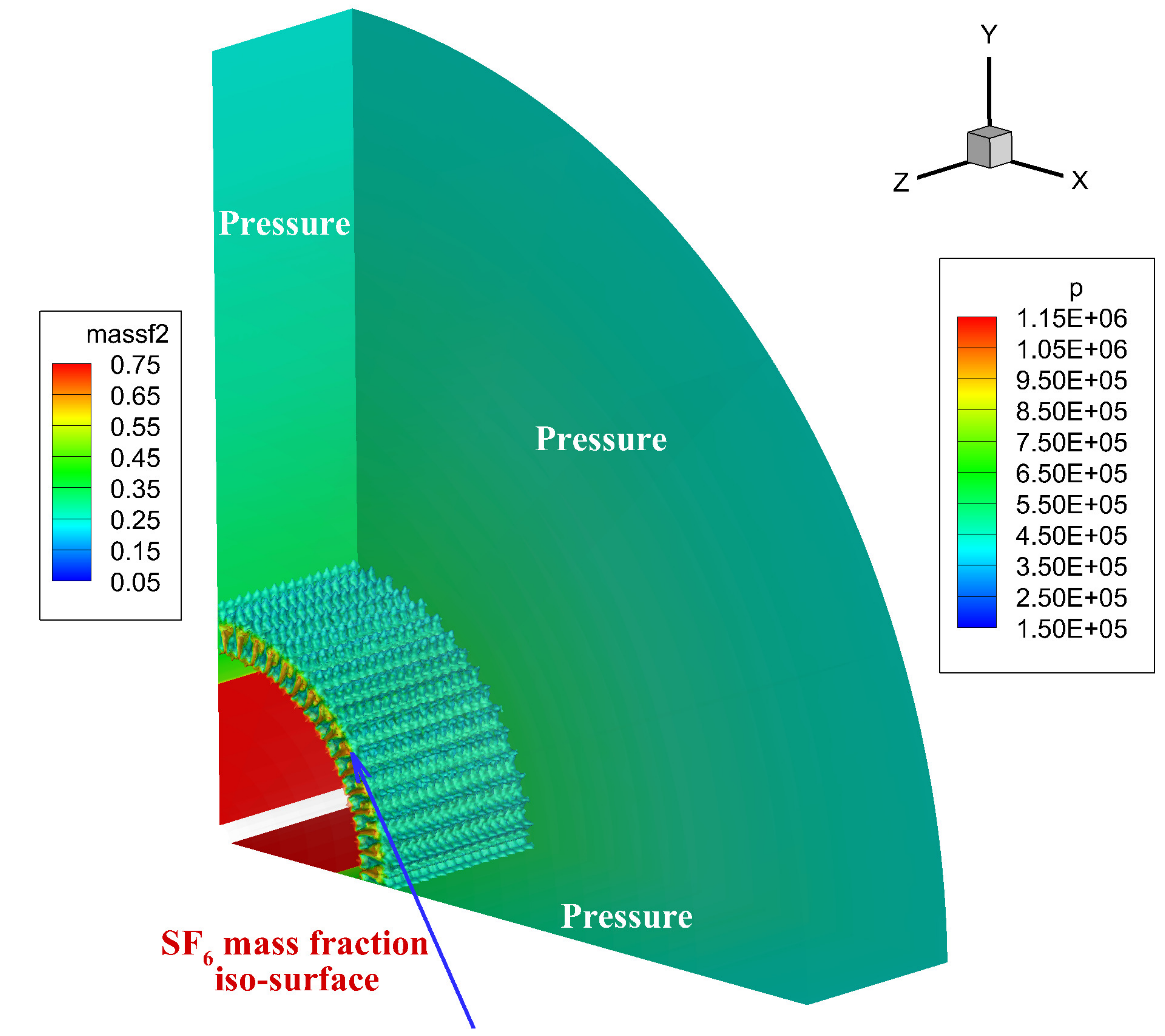}
\end{minipage}%
}%

\subfigure[$t=$1.740\ $ms$]{
\begin{minipage}[t]{0.45\linewidth}
\centering
\includegraphics[width=0.85\linewidth]{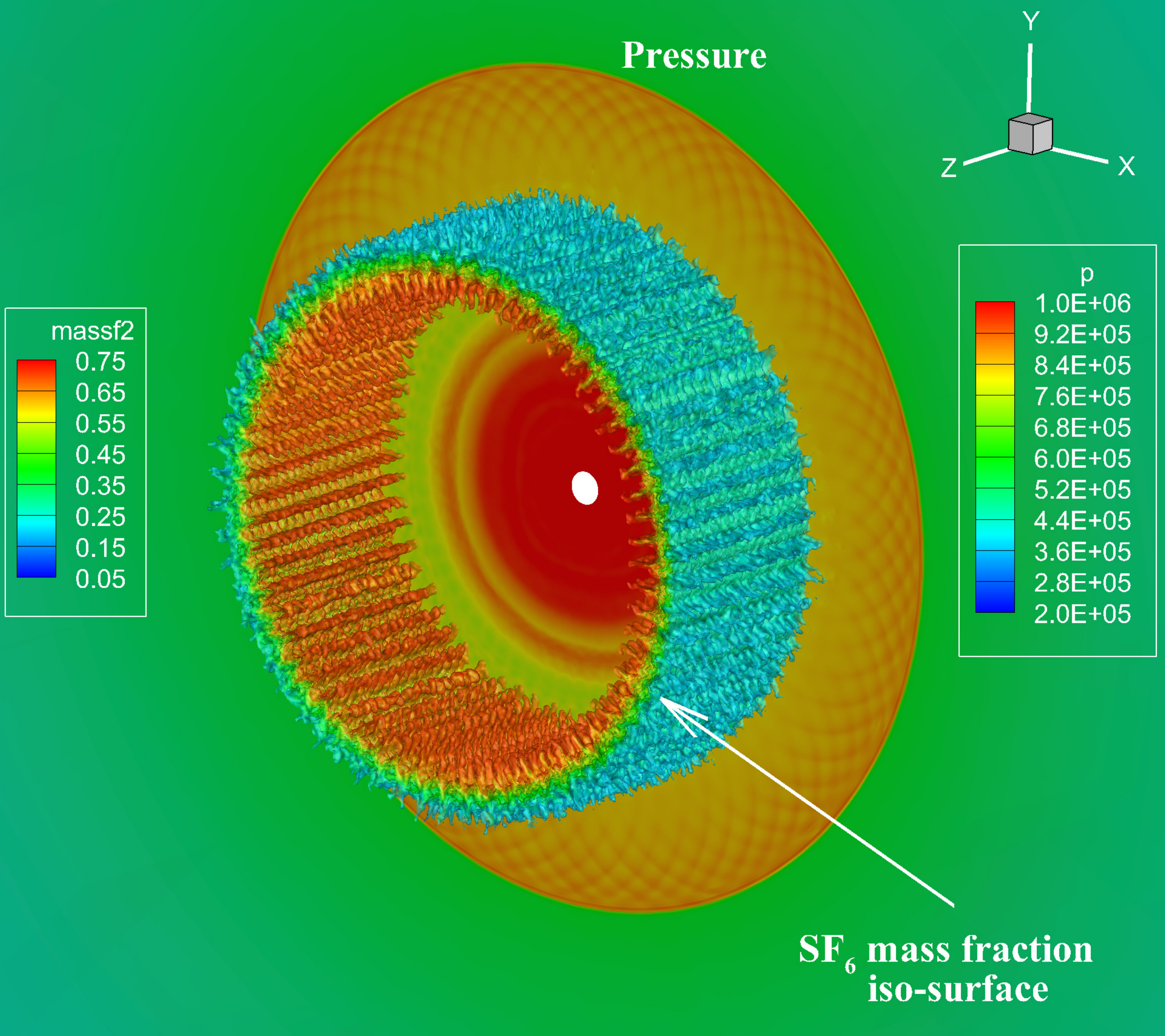}
\end{minipage}%
}%
\subfigure[$t=$2.040\ $ms$]{
\begin{minipage}[t]{0.45\linewidth}
\centering
\includegraphics[width=0.85\linewidth]{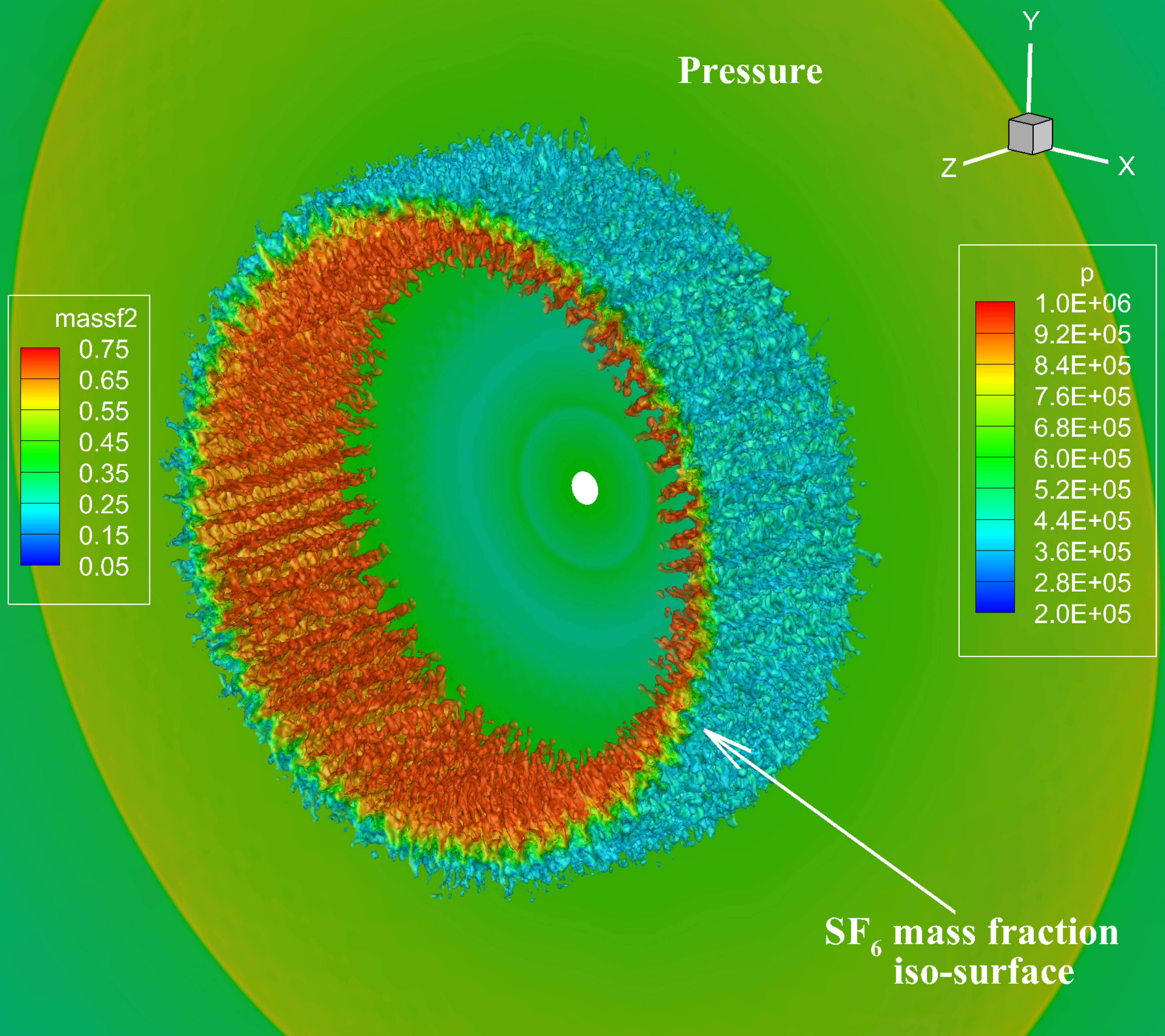}
\end{minipage}%
}%
\caption{\label{fig:fig8}  Evolutions of flow structures and shock waves propagations for Case 1.}
\end{figure*}

\begin{figure*}
\centering
\subfigure[$t=$0.840\ $ms$]{
\begin{minipage}[t]{0.5\linewidth}
\centering
\includegraphics[width=0.85\linewidth]{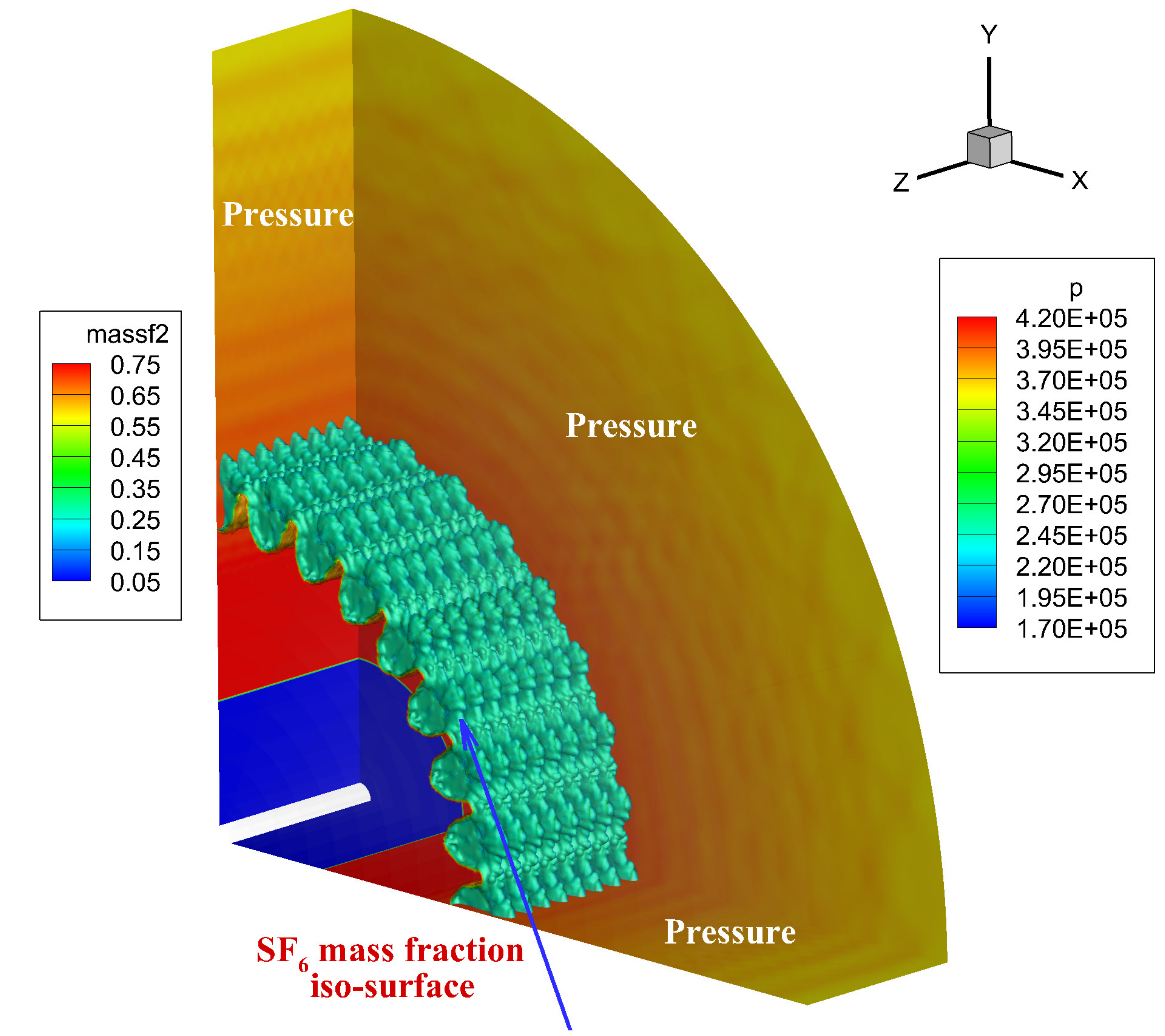}
\end{minipage}%
}%
\subfigure[$t=$1.440\ $ms$]{
\begin{minipage}[t]{0.5\linewidth}
\centering
\includegraphics[width=0.85\linewidth]{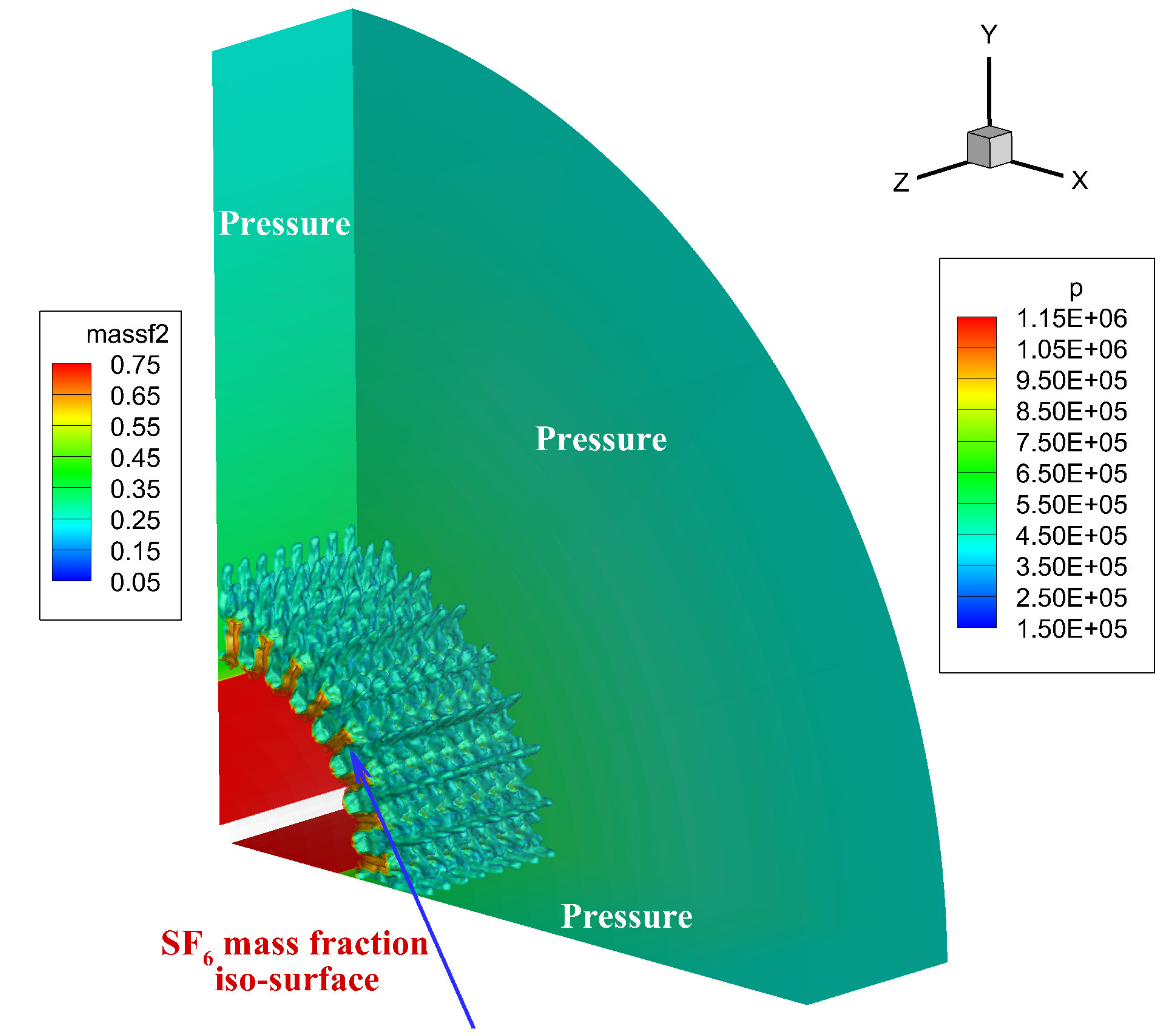}
\end{minipage}%
}%

\subfigure[$t=$1.740\ $ms$]{
\begin{minipage}[t]{0.45\linewidth}
\centering
\includegraphics[width=0.85\linewidth]{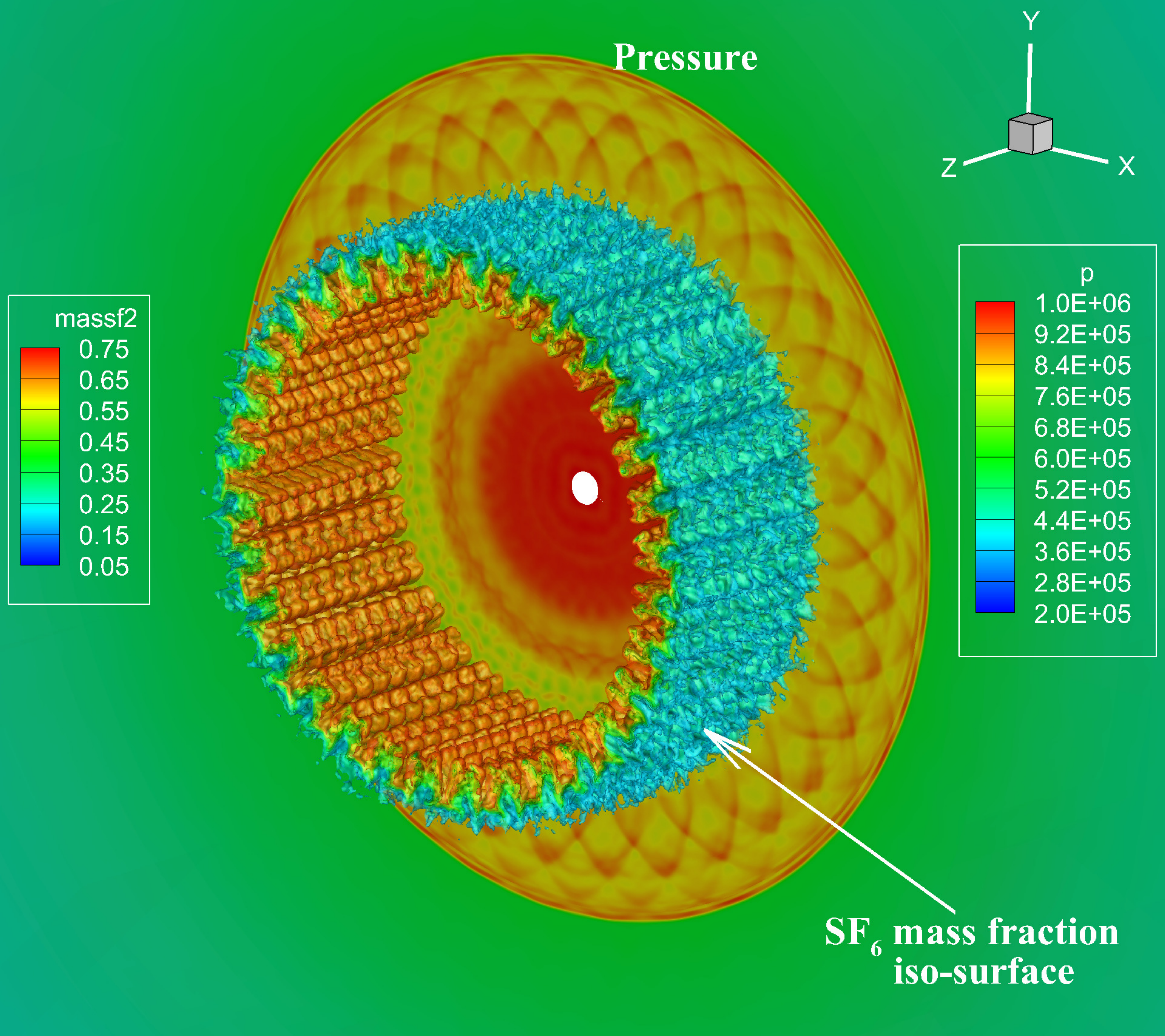}
\end{minipage}%
}%
\subfigure[$t=$2.040\ $ms$]{
\begin{minipage}[t]{0.45\linewidth}
\centering
\includegraphics[width=0.85\linewidth]{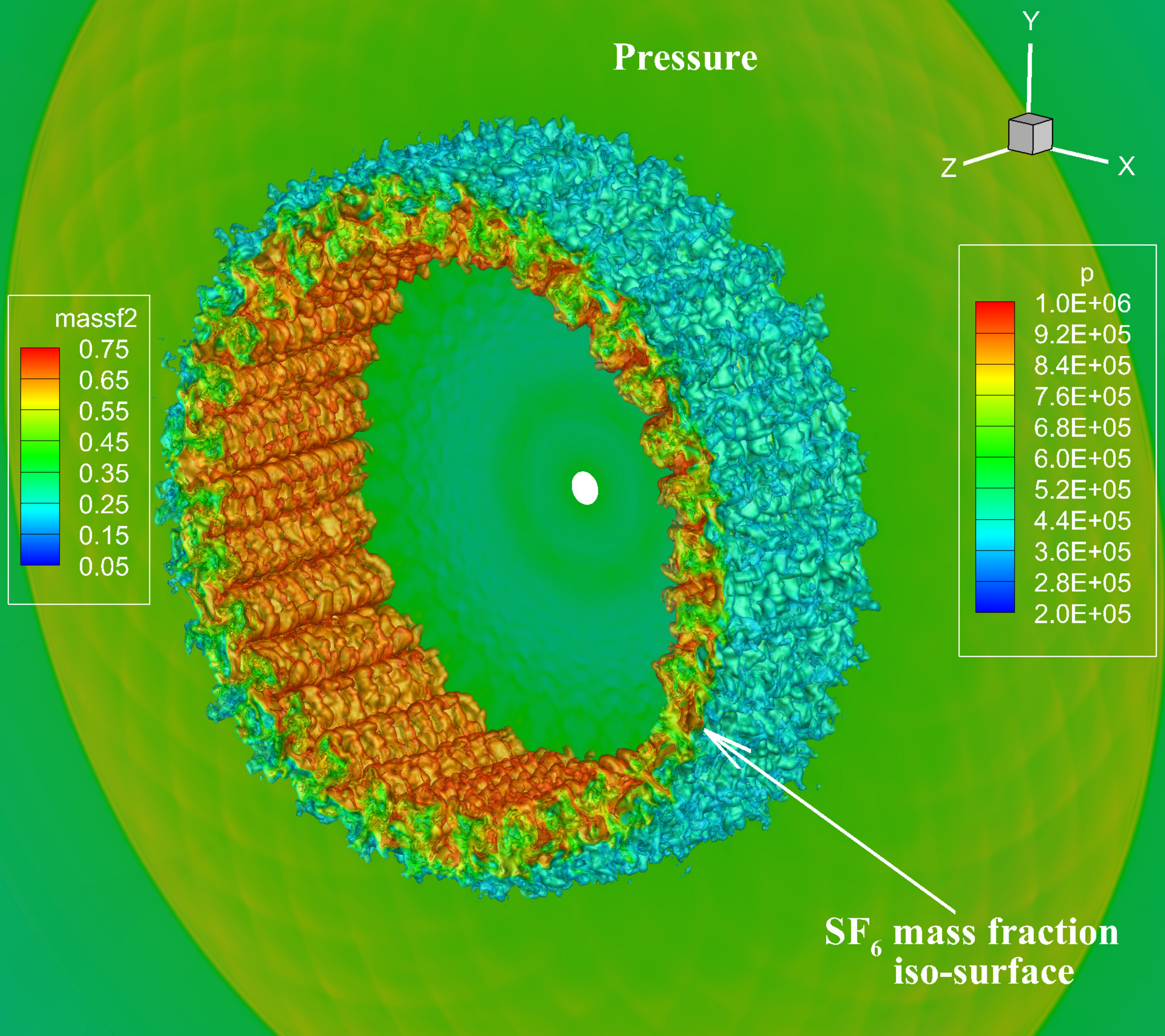}
\end{minipage}%
}%

\caption{\label{fig:fig9} Evolutions of flow structures and shock waves propagations for Case 2.}
\end{figure*}

\begin{figure*}
\centering

\subfigure[Case 1]{
\begin{minipage}[t]{0.45\linewidth}
\centering
\includegraphics[width=0.9\linewidth]{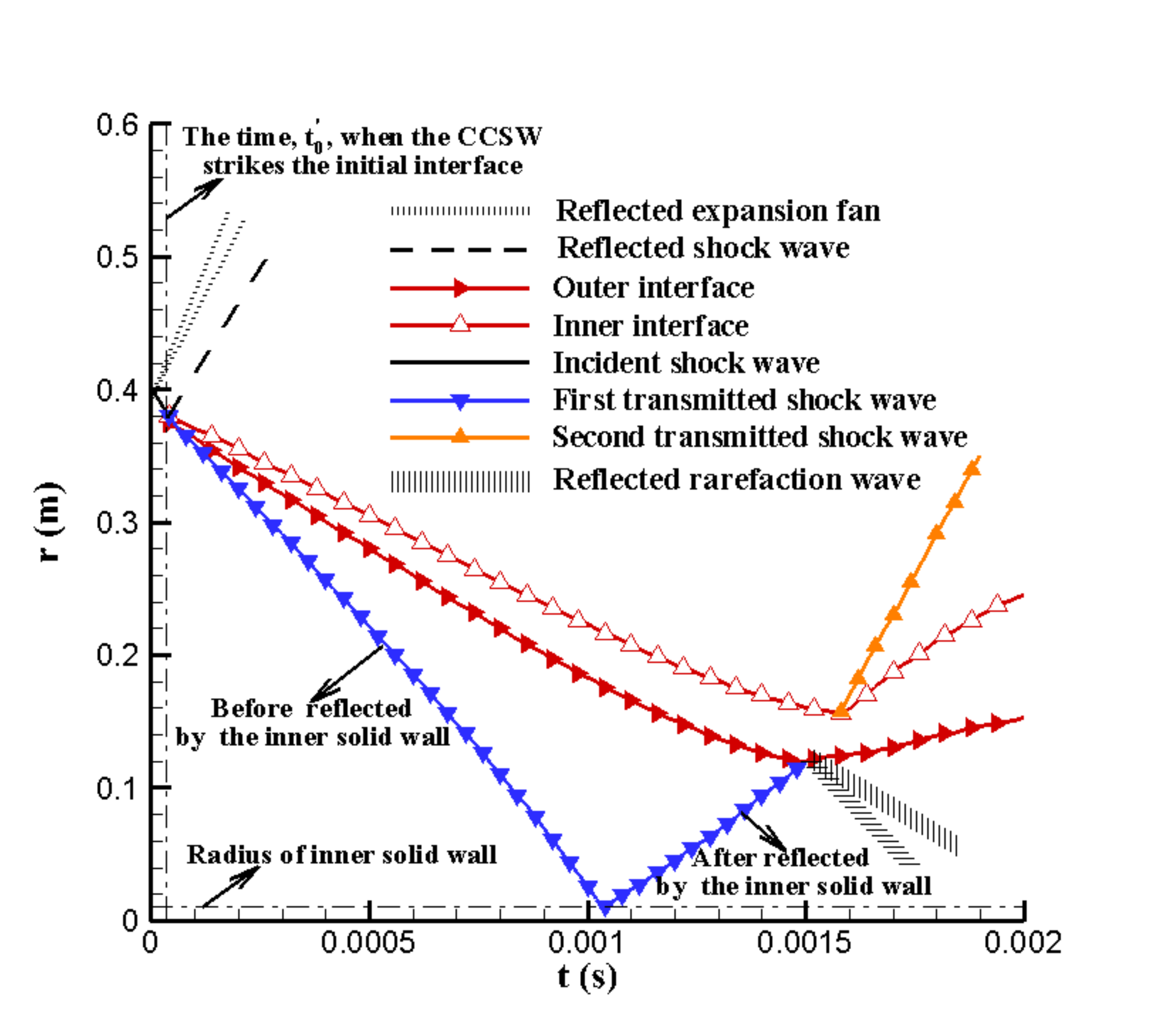}
\end{minipage}%
}%
\subfigure[Case 2]{
\begin{minipage}[t]{0.45\linewidth}
\centering
\includegraphics[width=0.9\linewidth]{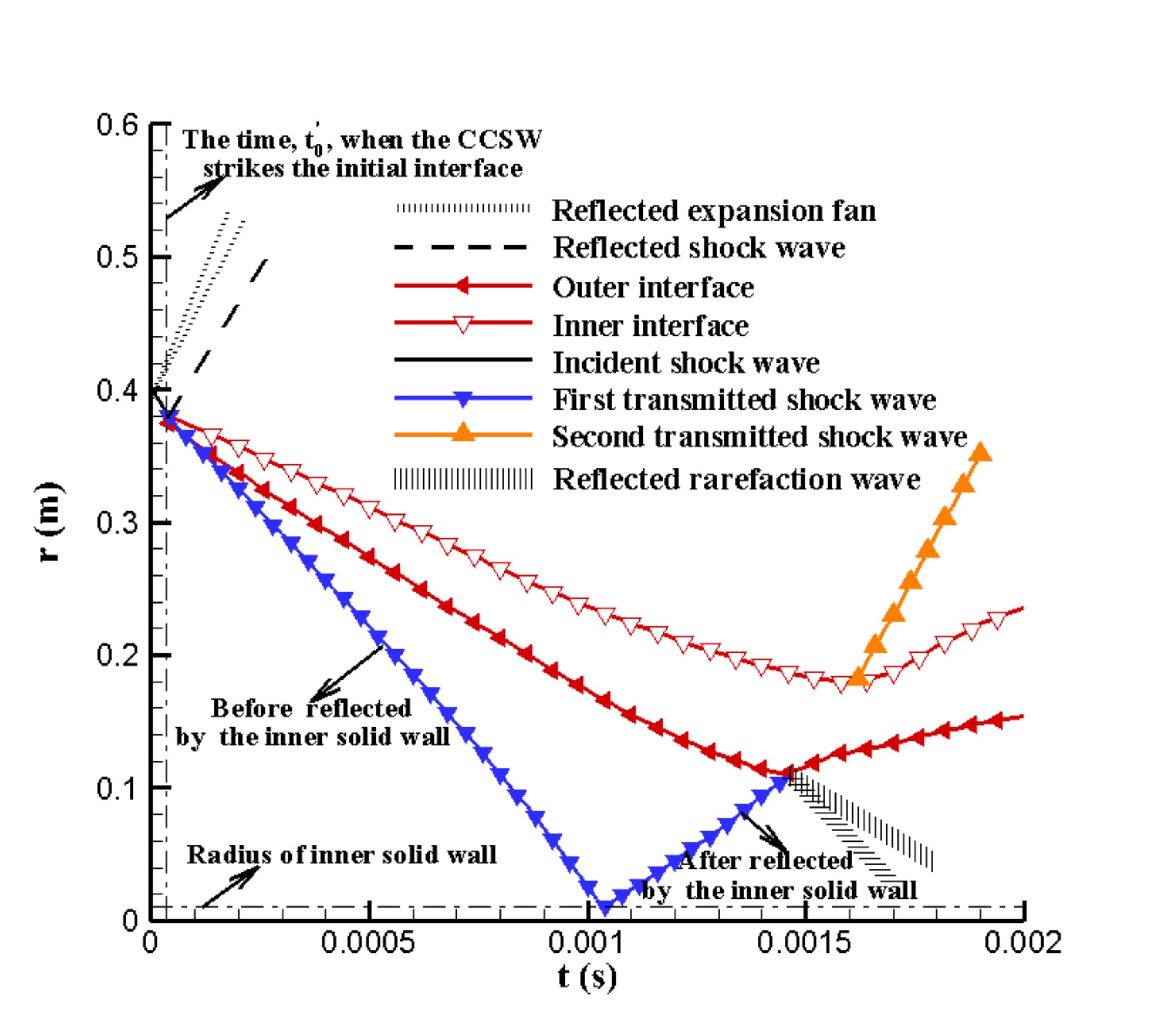}
\end{minipage}%
}%

\caption{\label{fig:fig10} Morphological wave patterns and quantitative evolutions of flow structures.}
\end{figure*}

\subsubsection{Scaling law of mixing width}
The self-similar scaling laws of mixing width for the RMI flow induced by planar shock wave are extensively investigated \cite{Thornber2010The, Tritschler2014On, Lombardini2012Transition, Thornber2017Late, thornber2011growth, thornber2012physics}. For the planar shock induced RMI flows without re-shock, Dimonte \cite{dimonte1995richtmyer} initially demonstrated that the mixing width is an exponential function versus time $\delta(t)\sim t^{\sigma}$. For the planar shock induced RMI flow with re-shock, later studies of Thornber
and Young \cite{thornber2011growth, thornber2012physics} show that the mixing width after re-shock scales as $\delta(t)\sim (t-t^{''}_{0})^{\sigma_r}$, where $t^{''}_{0}$ is a virtual time and always set to be the time of reshocking instant. Moreover, the exponents ${\sigma}$ and $\sigma_r$ for the above scaling laws would be varied depending on some factors such as Mach number of incident shock wave \cite{Lombardini2012Transition}. 
Recently, the evolutions of mixing width for the RMI flows induced by converging shock wave are paid more attention \cite{Lombardini2014Turbulent_a, Rafei2019three, boureima2018properties}, while the corresponding scaling laws are not fully reported and remain to be open issues. Consequently, in this part, we try to gain further insights into the scaling laws of mixing width for RMI flows induced by CCSW.

In the cylindrical coordinate system, the mixing width for RMI flows induced by CCSW can be defined as
\begin{equation}
\delta(t)=4\int_{r_{min}^{IMZ}}^{r_{max}^{IMZ}}<Y_{\text{SF}_6}>_{\varphi z}(1-<Y_{\text{SF}_6}>_{\varphi z})dr.
\end{equation}
In the above formulation, $r_{min}^{IMZ}$ and $r_{max}^{IMZ}$ are, respectively, the minimum radius and maximum radius of the inner mixing zone (IMZ) in which the average mass fraction of sulphur hexafluoride $<Y_{\text{SF}_6}>_{\varphi z} \in [0.05,0.70]$. Additionally, for arbitrary scalar $\phi$, $<\phi>_{\varphi z}$ is the ensemble average of $\phi$ on the cylindrical shell (in $\varphi z$ plane), which is defined as
\begin{equation}
<\phi>_{\varphi z}(r,t)=\frac{1}{2 \pi L_z}\iint \phi(r,\varphi,z,t)d{\varphi}dz.
\end{equation}

The evolutions of mixing width versus time for both cases and their corresponding evolutions on Log-Log scale are shown in Fig.11(a) and Fig.11(b), respectively. As depicted in Fig.(9), the evolutions of mixing width for both RMI flows induced by CCSW almost follow the same scaling laws. At early stage, the scaling law of mixing width is $\delta(t)\sim t^{0.65}$, which is quite similar to the scaling law of mixing width for planar shock driven RMI flow with 3D broadband perturbations on an initial interface \citep{Thornber2010The}. Additionally, at the earlier stage after re-shock, the mixing width for both cases scales as $\delta(t)\sim (t-t^{''}_{0})^{0.25}$ (with re-shocking instant $t^{''}_{0}\approx 1.5ms$). Actually, this scaling law of mixing width for CCSW induced RMI flows is also widely reported for RMI flows driven by planar shock during the earlier stage after re-shock \cite{zeng2018turbulent,Tritschler2014On, thornber2011growth}. However, at the later stage after re-shock for present two CCSW driven RMI flows, the mixing width seems to scale as $\delta(t)\sim t^{\sigma}$ again, while the scaling exponent becomes ${\sigma}\approx 1.75$. This recovering scaling law of $\delta(t)\sim t^{\sigma}$ at later stage after re-shock seems to be unique for CCSW driven RMI flows since no other similar results have been reported for planar shock driven RMI flows. It should be noted that, although we witness the same scaling laws of mixing width for the present two CCSW induced RMI flows, the scaling exponents at corresponding stages would be varied for other cases since they would largely depend on the mode of initial perturbations, impulsive Mach numbers and other factors \cite{Lombardini2012Transition}.  
\begin{figure}
\centering

\subfigure[Normal evolutions versus time]{
\begin{minipage}[t]{1.0\linewidth}
\centering
\includegraphics[width=1.0\linewidth]{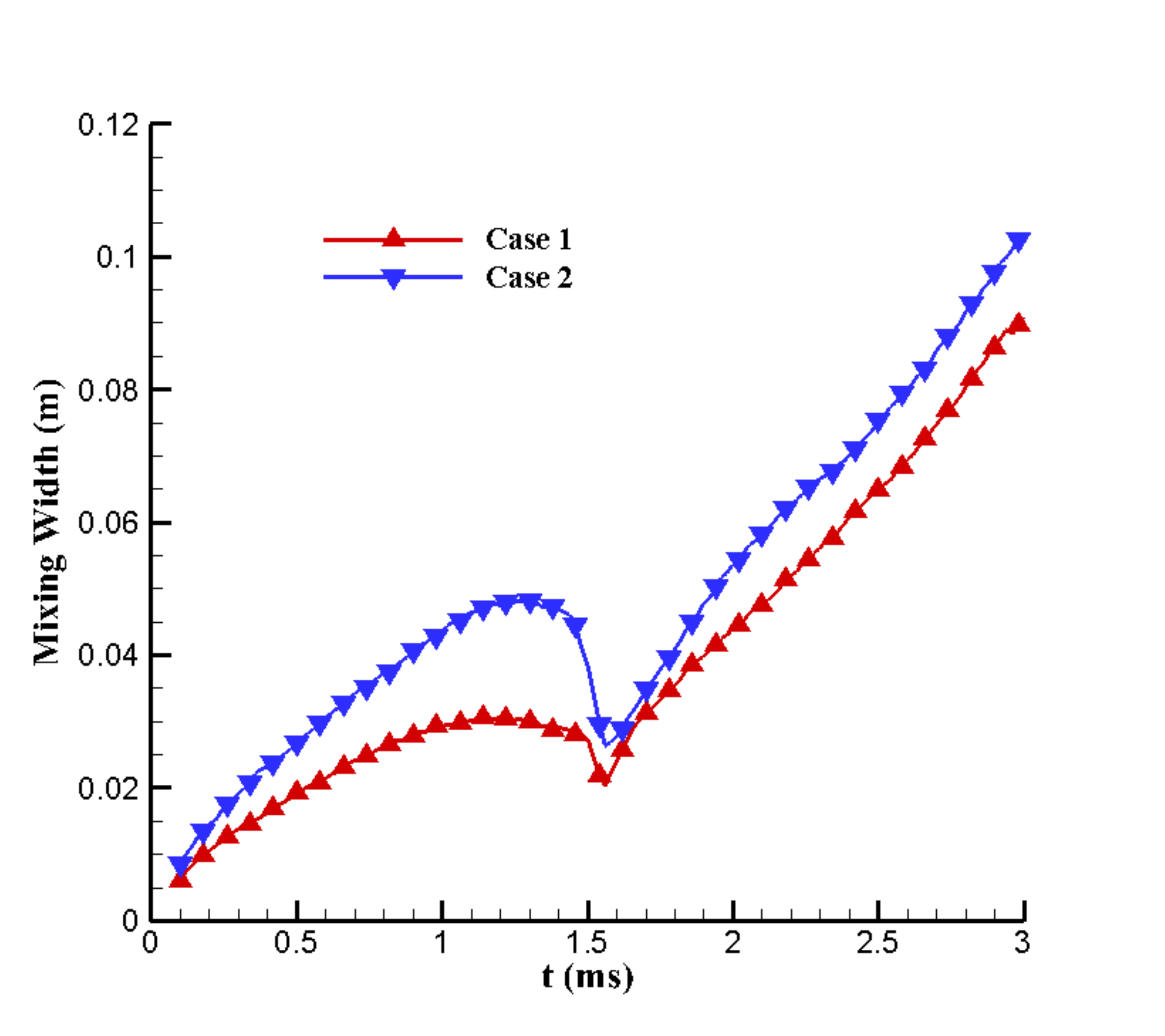}
\end{minipage}%
}%

\subfigure[Evolutions versus time on a Log-Log scale]{
\begin{minipage}[t]{1.0\linewidth}
\centering
\includegraphics[width=1.0\linewidth]{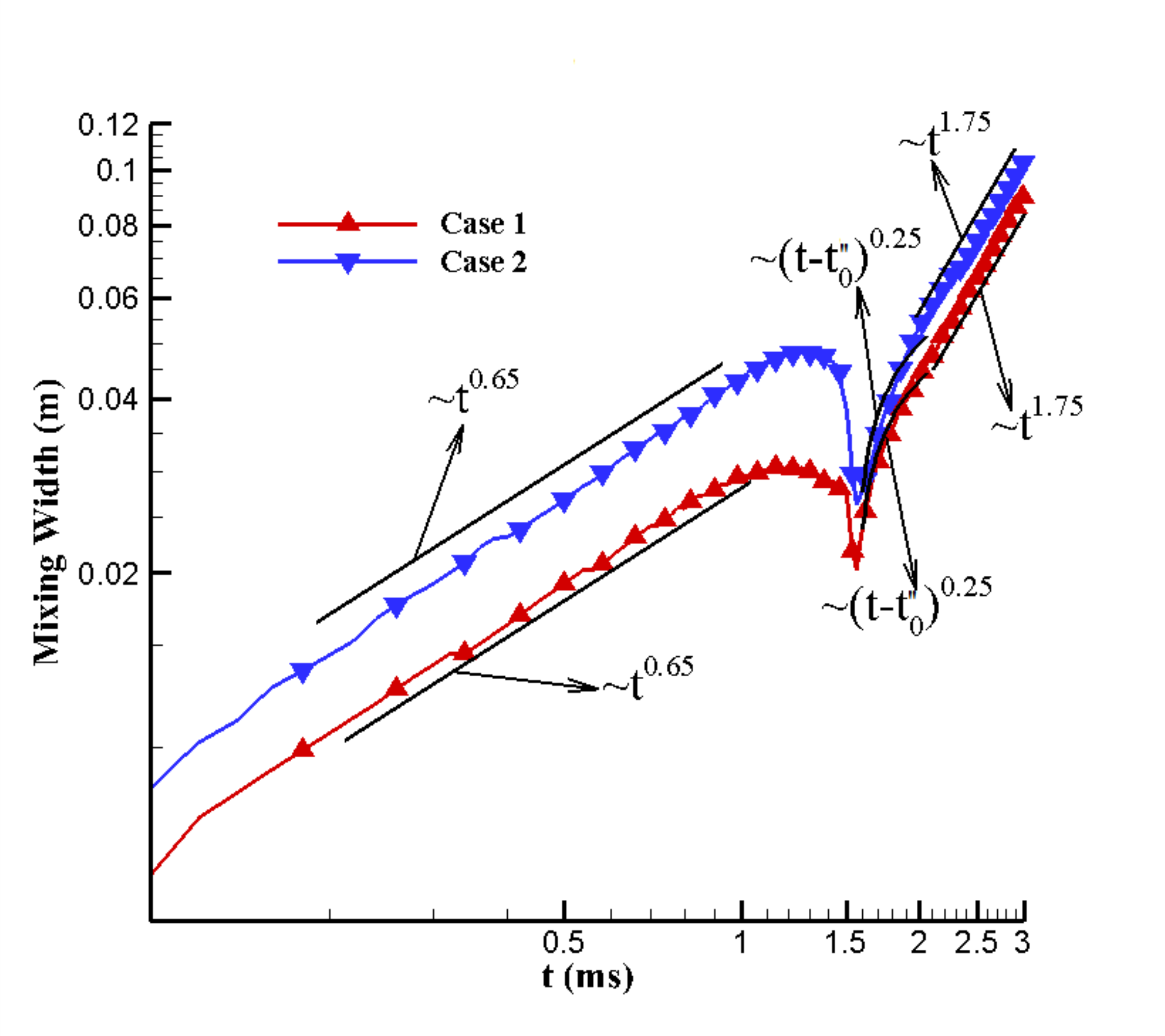}
\end{minipage}%
}%

\caption{\label{fig:fig11} Temporal evolutions of mixing width.}
\end{figure}

\subsubsection{Temporal asymptotic behaviors of mixing parameters}
As mentioned above, although the flow structures are characterized by the growth of perturbation amplitude before re-shock, the fluids' mixing is dramatically enhanced after re-shocked. To figure out the level of fluids’ mixing as well as the isotropy/homogeneity properties of the mixing zone, we quantitatively investigate the temporal asymptotic behaviors of molecular mixing fraction $\Theta$, local anisotropy $a_{i,vol}$ and density-specific volume correlation $b_{vol}$ in this subsection.

The molecular mixing fraction can characterize the relative amount of molecularly mixed fluid within the mixing layer. It can be interpreted as the ratio of molecular mixing to large-scale entrainment by convection motion. Following the definition of Youngs \cite{youngs1991three,youngs1994numerical}, in the cylindrical coordinate system, the formulation of molecular mixing fraction can be expressed as

\begin{equation}
\Theta(t)=\frac{\int_{r_{min}^{IMZ}}^{r_{max}^{IMZ}}<Y_{\text{SF}_6} (1-Y_{\text{SF}_6})>_{\varphi z}dr}{\int_{r_{min}^{IMZ}}^{r_{max}^{IMZ}}<Y_{\text{SF}_6}>_{\varphi z}(1-<Y_{\text{SF}_6}>_{\varphi z})dr}.
\end{equation}

The temporal asymptotic behaviors of the molecular mixing fraction for present two CCSW induced RMI flows are shown in Fig.12. As shown in Fig.12, the ratios of molecular mixing to large-scale entrainment by convection motion are relatively small for both cases at the early stage before re-shock, while they increase as the instabilities evolve. Additionally, the molecular mixing between fluids for both cases is sharply increased, to some extent, after the gases interfaces are re-shocked by the reflected FTSW (after $t \approx 1.5\times 10^{-3} s$). Moreover, at the later stage after re-shock, the evolutions of molecular mixing fraction for both cases become asymptotic, with a final value being 0.93 approximately. Actually, the above asymptotic behavior of molecular mixing fraction for the present CCSW induced RMI flows highly resembles that for some planar shock driven RMI flows which are well reported numerically and experimentally \cite{zeng2018turbulent, orlicz2013incident} 
\begin{figure}
\centering
\includegraphics[width=1.0\linewidth]{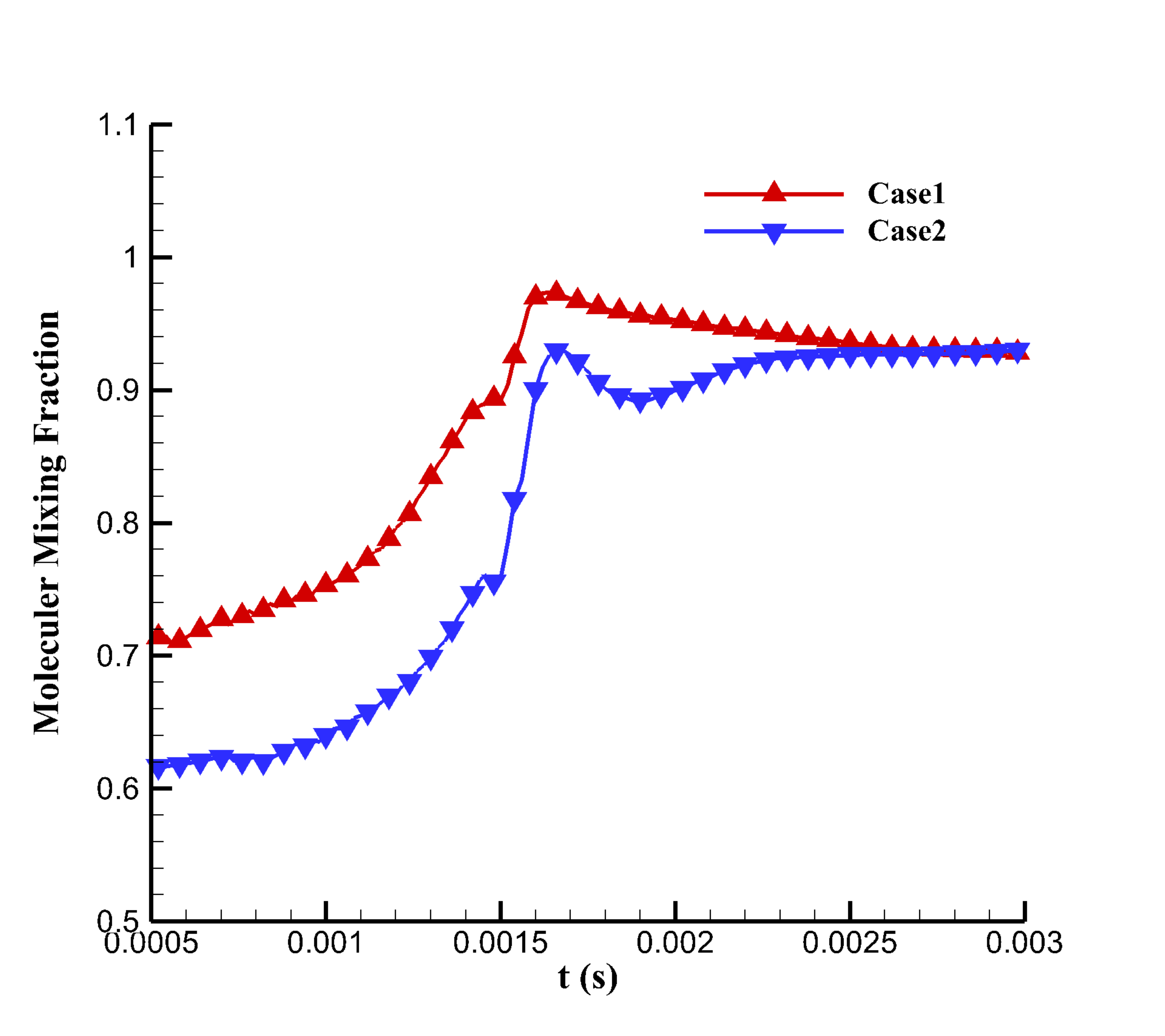}
\caption{\label{fig:fig12} Temporal evolutions of the molecular mixing fraction.}
\end{figure}

For mixing flows, the anisotropy and inhomogeneity are of significance since both of them are important to large-eddy and Reynolds-averaged Navier-Stokes modeling \cite{cabot2013statistical}. To figure out the properties of local anisotropy and inhomogeneity of fluids’ mixing for the present two CCSW driven RMI flows, we investigate the temporal asymptotic behaviors of the volume-averaged anisotropy $a_{i,vol}$ and the volume-averaged density-specific volume correlation $b_{vol}$, respectively. Their formulations are given as follows
\begin{equation}
a_{i,vol}=\frac{1}{\Delta r}\int_{r_{min}^{IMZ}}^{r_{max}^{IMZ}} <\frac{ \left | u_i^{''} \right |}{\left | u_r^{''} \right |+\left | u_{\varphi}^{''} \right |+\left | u_z^{''} \right |}-\frac{1}{3}>_{\varphi z}dr,
\end{equation}
\begin{equation}
b_{vol}=\frac{1}{\Delta r}\int_{r_{min}^{IMZ}}^{r_{max}^{IMZ}}(<\frac{1}{\rho}>_{\varphi z}<\rho>_{\varphi z}-1)dr.
\end{equation}
In the above two equations, $\Delta r= r_{max}^{IMZ}-r_{min}^{IMZ}$ is
the length of the inner mixing zone in the radial direction. Additionally, for arbitrary scalar $\phi$, its fluctuating part, $\phi^{''}$, is given by
\begin{equation}
\phi^{''}=\phi-\overline{\phi},
\end{equation}
where $\overline{\phi}=<\rho \phi>_{\varphi z}/<\rho>_{\varphi z}$ is the ensemble Favre average of $\phi$ on the cylindrical shell. Moreover, for $i=r,\phi$ and $z$, $u_i^{''}$ respectively indicates the fluctuating part of radial, azimuthal and axial velocity, and, $a_{i,vol}$ respectively denotes the corresponding volume-averaged anisotropy in radial, azimuthal and axial direction. 

Based on Eq.(26), the volume-averaged anisotropy $a_{i,vol}$ would range from $-\frac{1}{3}$ to $\frac{2}{3}$, manifesting the ratio of the TKE in a specific direction to the total TKE. Larger value of $a_{i,vol}$ implies larger fluctuation or TKE in the corresponding direction. Moreover, the positive value of $a_{i,vol}$ indicates that the TKE in the corresponding direction is dominant, while the negative value of $a_{i,vol}$ implies that the TKE in the corresponding direction is less important. For isotropically mixing flow, $a_{i,vol}$ should be almost zero in all directions. Fig.13 shows the temporal evolutions of the volume-averaged anisotropy in all three directions for both cases. As shown in Fig.13, the volume-averaged anisotropy in each direction would achieve a final asymptotic value for both case. However, the temporal asymptotic behavior of $a_{i,vol}$ for the present CCSW induced RMI flows is different from that for planar shock driven RMI flows. For the planar shock driven RMI flows, the magnitude of $a_{i,vol}$ in all directions would achieve a very small asymptotic value at later stage \cite{zeng2018turbulent, orlicz2013incident}, implying that the flows would become much less anisotropic. However, for the present CCSW driven RMI flows during the duration of simulations, the temporal asymptotic values in three directions follow the law of $a_{r,vol}>0>a_{\varphi,vol}>a_{z,vol}$ and the magnitudes of $a_{r,vol}$ and $a_{z,vol}$ are much larger than zero to some extent. These results indicate that, during the duration of simulations, the fluids mixing would always be anisotropic for the present CCSW driven RMI flows. Actually, the above results are consistent with the corresponding evolutions of TKE. For CCSW driven RMI flows, the TKE in the radial direction and the TKE in the azimuthal direction are always more important than the TKE in the axial direction, simply because the converging and expanding effects will continuously perturb the flow field in these two directions.
     
\begin{figure}
\centering
\includegraphics[width=1.0\linewidth]{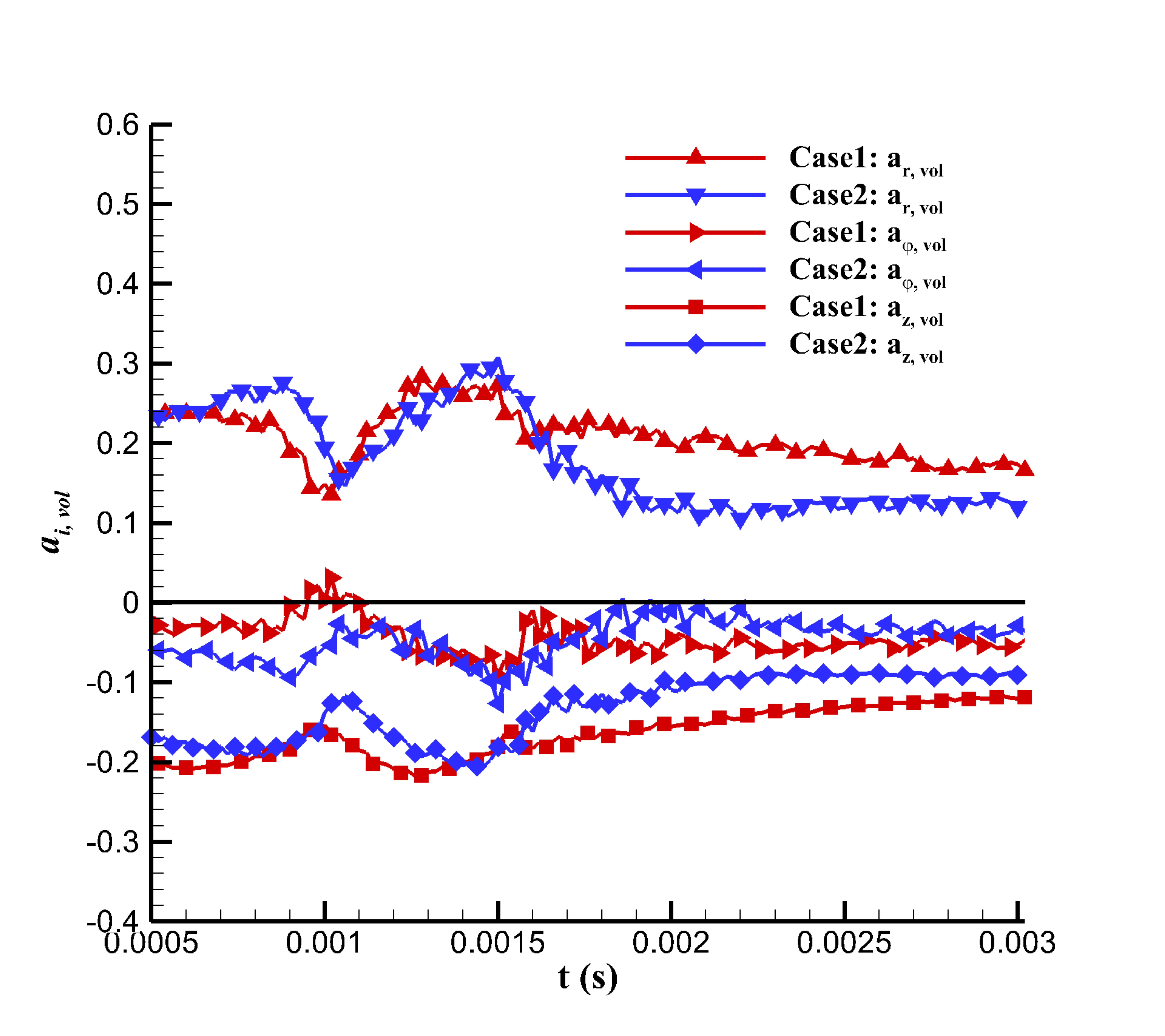}
\caption{\label{fig:fig13} Temporal evolutions of the anisotropy.}
\end{figure}

The volume-averaged density-specific volume correlation is critical in second-moment turbulence modeling for variable density flows \cite{livescu2009high}. According to Eq.(27),  $b_{vol}$ is a non-negative parameter. For nearly homogeneous flow, $b_{vol}$ would become very small. However, if the fluids' mixing is spatially inhomogeneous, the value of $b_{vol}$ would be large. Fig.14 shows the temporal evolutions of the volume-averaged density-specific volume correlation for the present two CCSW induced RMI flows. As shown in Fig.14, the volume-averaged density-specific volume correlation for both cases is decreasing on the whole. Moreover, at the later stage after re-shock, the volume-averaged density-specific volume correlation for both cases would asymptotically achieve the same relatively small value of 0.04. These results imply that the fluids' mixing for the present CCSW driven flows would become much less inhomogeneous at later stage after re-shock.

\begin{figure}
\centering
\includegraphics[width=1.0\linewidth]{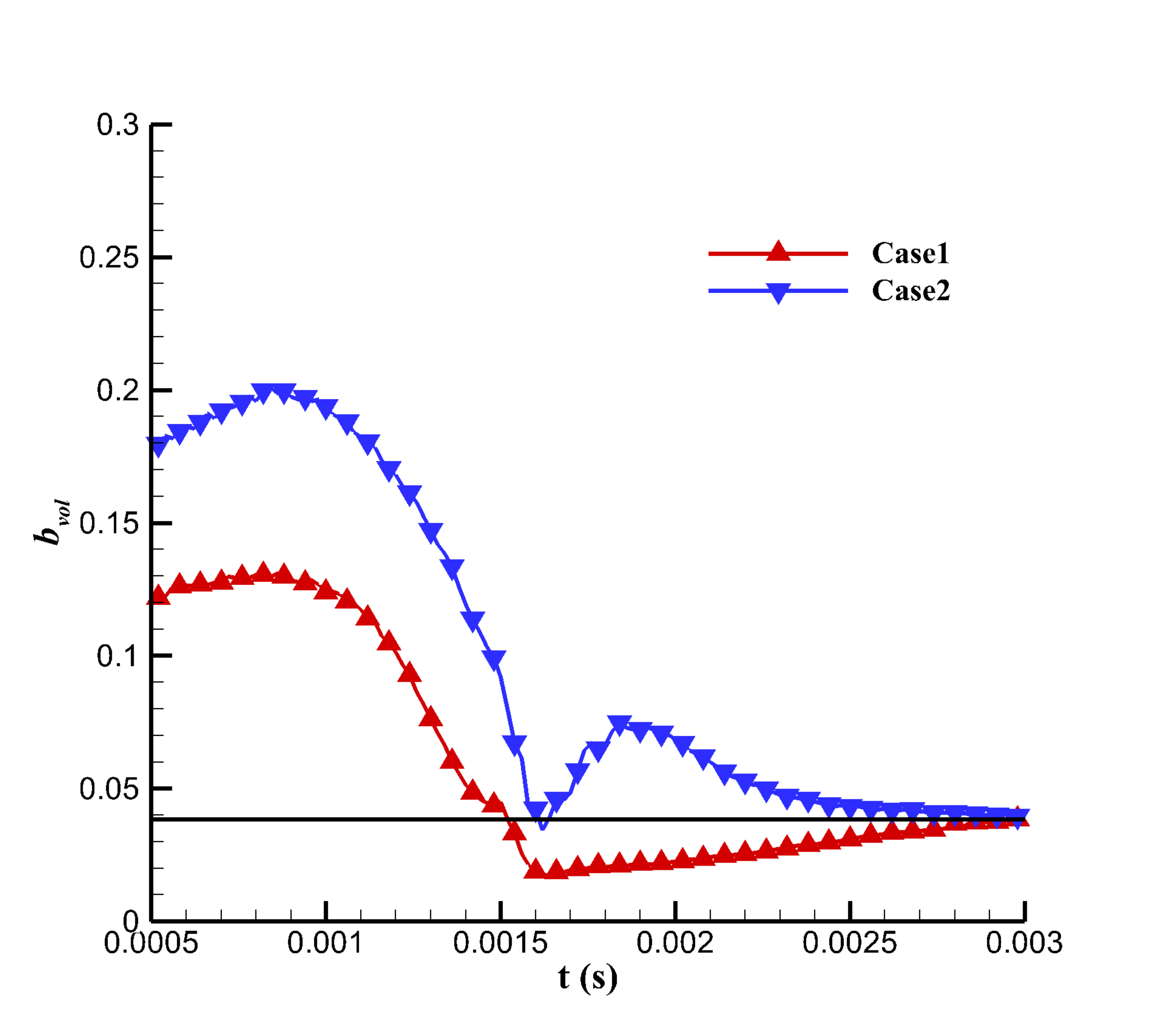}
\caption{\label{fig:fig14} Temporal evolutions of the density-specific volume correlation.}
\end{figure}

\subsubsection{Turbulent kinetic energy spectrums}

As mentioned by Tritschler \cite{Tritschler2014On}, a fully isotropic mixing zone is never obtained for shock induced fluids' mixing flows, although the fluid's mixing would become less anisotropy and less inhomogeneous at later stage. However, the theory of TKE spectrum for homogeneous isotropic turbulence is often used as the theoretical framework for the numerical analyses of shock induced RMI flows \cite{Tritschler2014On}. According to this theory, there would be a broadened inertial range in TKE spectrums once the fluids' mixing is enhanced. Therefore, alternative to the aforementioned analyses in physical space, the enhanced fluids' mixing after re-shock is analyzed in Fourier representation for both cases in this subsection. Additionally, due to the moderately small cell number in the axial direction, the 2D TKE spectrums analyses in $\varphi z$ plane would conceal the characteristics of TKE spectrums at relatively high wave numbers in the azimuthal direction (which would be more important for the present CCSW induced RMI flows since, as mentioned above, the TKE in the azimuthal direction would be more important to some extent). Consequently, only the TKE spectrums in the azimuthal direction after re-shock are analyzed for both cases in this subsection.

In the cylindrical coordinate system, the average TKE spectrum for the inner mixing zone in the azimuthal direction is given by 
\begin{equation}
E^a(k_\varphi, t)=\frac{1}{\Delta r}\int_{r_{min}^{IMZ}}^{r_{max}^{IMZ}}<E(r, k_\varphi, z,t)>_{\varphi z}dr,
\end{equation}
where $E(r, k_\varphi, z,t)$ is the TKE spectrum of wave number $k_\varphi$ in azimuthal direction at time $t$ on specific radial $r$ and axial position $z$. The formulation of TKE spectrum is given by 
\begin{equation}
E(r, k_\varphi, z,t)=\widehat{u_{r}^{''}}\widehat{u_{r}^{''}}^*+\widehat{u_{\varphi}^{''}}\widehat{u_{\varphi}^{''}}^*+\widehat{u_{z}^{''}}\widehat{u_{z}^{''}}^*.
\end{equation}
In the above equation, for arbitrary scalar $\phi$, $\widehat{\phi }$ denotes its Fourier transform in the azimuthal direction, and $\widehat{\phi }^*$ indicates the corresponding complex conjugate of $\widehat{\phi }$. Fig.15 shows the average TKE spectrums at three instants of the later stage after re-shock for the present two CCSW driven RMI flows. As depicted in Fig.15, at  the relatively early stage after re-shock $(t=2.440 ms)$, the decaying law of $k^{-5/3}$ for the TKE spectrums of these two cases only locates in a narrowband of wave numbers. However, as the flows evolve, the TKE spectrums for both cases would decay with a slope of $k^{-5/3}$ in a broader range of wave numbers. These results imply that the inertial range is extended during the developing process and manifest, to some extent, that the fluids' mixing is enhanced at later time after re-shock.
\begin{figure}
\centering
\subfigure[Case 1]{
\begin{minipage}[t]{1.0\linewidth}
\centering
\includegraphics[width=1.0\linewidth]{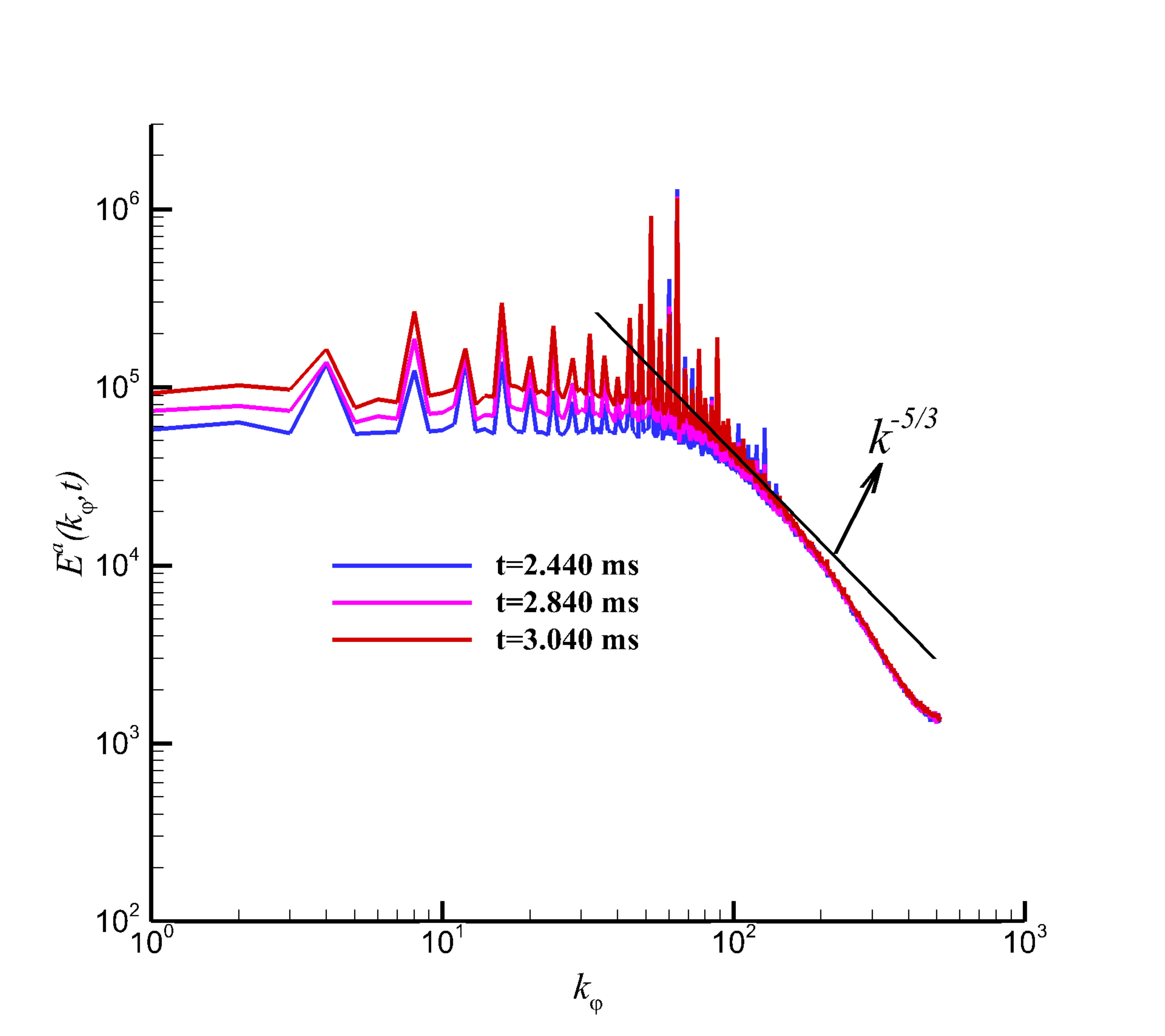}
\end{minipage}%
}%

\subfigure[Case 2]{
\begin{minipage}[t]{1.0\linewidth}
\centering
\includegraphics[width=1.0\linewidth]{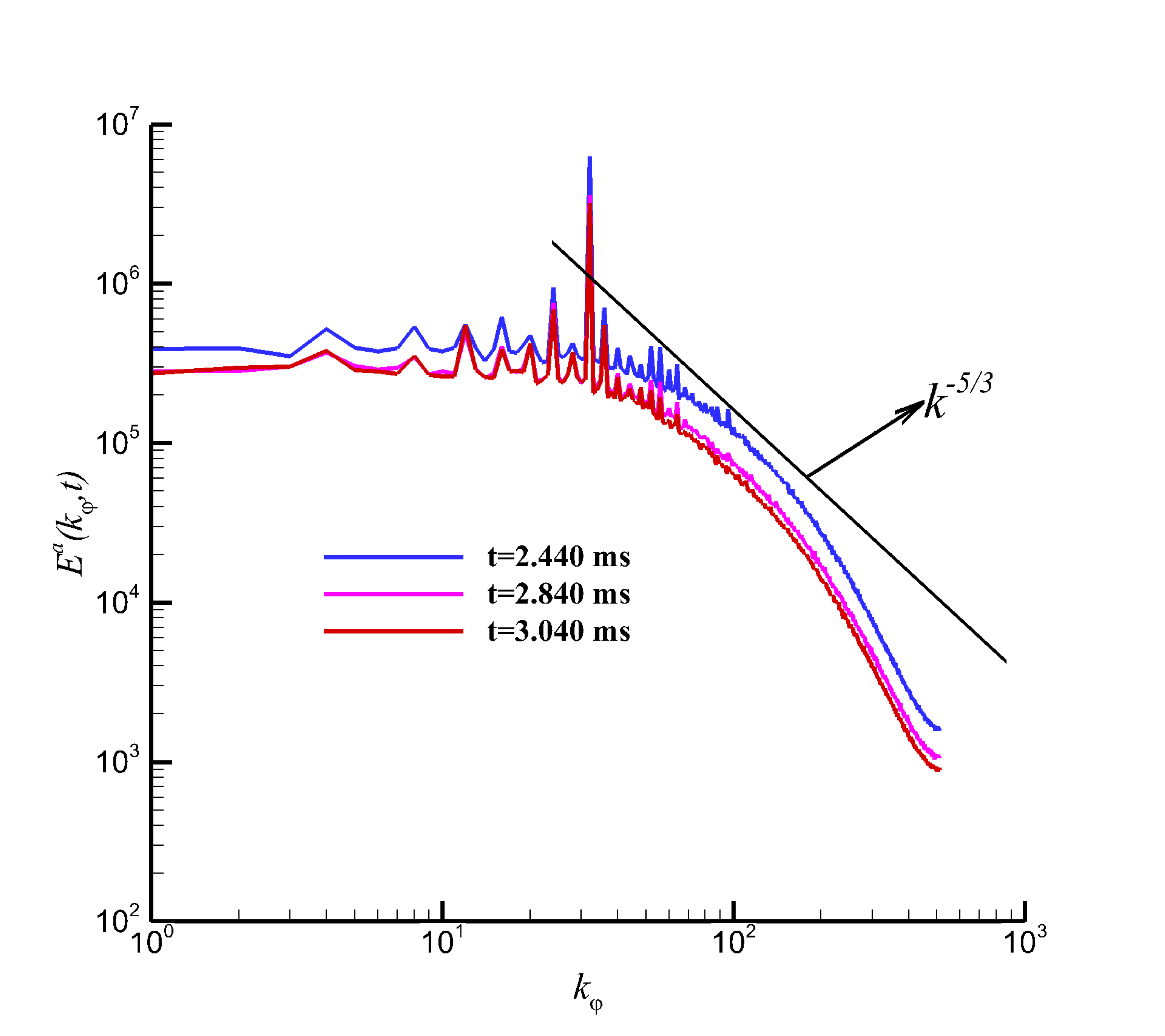}
\end{minipage}%
}%

\caption{\label{fig:fig15} Average TKE spectrums in the azimuthal direction.}
\end{figure}

\section{CONCLUSIONS}
Based on the theory of shock tube, we successfully generate "pure" CCSW without a following contact surface. Additionally, studies on the amplitude growth of gases interfaces manifest that the generated CCSW is efficient for studying on CCSW induced RMI flows. Then, the instabilities and fluids’ mixing behaviors of two gases interfaces driven by CCSW are numerically investigated using high resolution FV method. The morphological wave patterns and the evolutions of flow structures imply that the instabilities of the interfaces are characterized by the growth of perturbation amplitude before re-shock, while the fluids' mixing is dramatically enhanced after re-shock. Detailed analyses of the fluids' mixing parameters show that the evolutions of mixing width and other mixing parameters could achieve the same laws of temporal behavior for the present two CCSW induced RMI flows, which indicates the existences of scaling law and temporal asymptotic behaviors for fluids' mixing parameters in the mixing zone. 
Additionally, due to the converging/expanding effects on the flow fields, the motions of shock wave and inner/outer gases interfaces for the present CCSW driven RMI flows are nonlinear versus time, which is quite different from the results of planar shock wave induced RMI flows. These nonlinear developments of flow fields would lead to some unique features for the fluids' mixing of CCSW driven RMI flows, one of which is that, at later stage after re-shock, the fluids' mixing would be less isotropic than that of planar shock wave induced RMI flows although both of them would reach final temporal asymptotic behaviors in each direction. Further analyses of TKE spectrums in the azimuthal direction at later stage after re-shock also witness the $k^{-5/3}$ decaying law of TKE spectrums for the present CCSW driven RMI flows. Both the temporal behaviors of mixing parameters and the decaying law of TKE spectrums manifest that the fluids’ mixing is indeed enhanced at later time after re-shock.

\section*{AVAILABILITY OF DATA}
Raw data were generated at the TH-2 Supercomputer. Derived data supporting the
findings of this study are available from the corresponding author upon reasonable request.

\begin{acknowledgments}
The work of this paper is supported by 2016YFA0401200 of the National Key Research and Development Program of China and U1430235 of the National Natural Science Foundation of China.
\end{acknowledgments} 

\nocite{*}

\bibliography{RMI_CCSW}
\end{document}